\documentclass[aps,preprint,nofootinbib,eqsecnum]{revtex4}%
\usepackage{amsmath}
\usepackage{amsfonts}
\usepackage{amssymb}
\usepackage{graphicx}
\usepackage{amscd}%
\allowdisplaybreaks
\begin{document}
\noindent{\small USC-05/HEP-B3\hfill\hfill hep-th/0512348 }

\bigskip

\begin{center}
{\Large \textbf{Twistor Transform in d Dimensions}}

{\Large \textbf{and a Unifying Role for Twistors}}

{\vskip0.8cm}

\textbf{Itzhak Bars$^{a}$ and Mois\'{e}s Pic\'{o}n$^{a,b}$}

\textit{{\ \smallskip}\\[0pt]$^{a)}$ Department of Physics and
Astronomy,\\[0pt]University of Southern California, Los Angeles, CA
90089-0484, USA \\[0pt]\noindent{\smallskip}\\[0pt]$^{b)}$ Departamento de
F\'{\i}sica Te\'{o}rica, \\[0pt]University of Valencia and IFIC (CSIC-UVEG),
46100-Burjasot (Valencia), Spain\\[0pt]\noindent{\ \smallskip}\\[0pt]\bigskip}

\textbf{Abstract}
\end{center}

Twistors in four dimensions $d=4$ have provided a convenient description of
massless particles with any spin, and this led to remarkable computational
techniques in Yang-Mills field theory. Recently it was shown that the same
$d=4$ twistor provides also a unified description of an assortment of other
particle dynamical systems, including special examples of massless or massive
particles, relativistic or non-relativistic, interacting or non-interacting,
in flat space or curved spaces. In this paper, using 2T-physics as the primary
theory, we derive the general twistor transform in $d$-dimensions that applies
to all cases, and show that these more general twistor transforms provide $d$
dimensional holographic images of an underlying phase space in flat spacetime
in $d+2$ dimensions. Certain parameters, such as mass, parameters of spacetime
metric, and some coupling constants appear as moduli in the holographic image
while projecting from $d+2$ dimensions to $\left(  d-1\right)  +1$ dimensions
or to twistors. We also extend the concept of twistors to include the phase
space of D-branes, and give the corresponding twistor transform. The unifying
role for the same twistor that describes an assortment of dynamical systems
persists in general, including D-branes. Except for a few special cases in low
dimensions that exist in the literature, our twistors are new.

\newpage

\section{Bottom$\rightarrow$up approach to spacetime}

In this section we will discuss mainly spinless particles to establish some
concepts in a familiar setting. A similar analysis applies to particles with
spin\footnote{A detailed discussion of the spinning particle and the related
generalized twistors, including higher dimensions and supersymmetry, will be
given in a future paper \cite{twistorsSpin}.}, therefore when we discuss
twistors in four dimensions, $d=4$ in section-\ref{twistd4}, we include the
spin in some of the discussion. Generalizations of the twistors to higher
dimensions, and D-branes, will be presented in the following sections. Other
generalizations, including spinning particles, supersymmetry, compactified
internal spaces, will be given elsewhere.

\subsection{Phase space}

A massless and spinless relativistic particle in $\left(  d-1\right)  $ space
dimensions is described by its position-momentum phase space coordinates
$\left(  \vec{x}^{i}\left(  t\right)  ,\vec{p}^{i}\left(  t\right)  \right)
,$ $i=1,2,\cdots,\left(  d-1\right)  $ while its time development is governed
by the Hamiltonian $H=\left\vert \vec{p}\right\vert =\sqrt{\vec{p}^{2}}.$
Hence its action is
\begin{equation}
S=\int dt\left(  \partial_{t}\vec{x}\cdot\vec{p}-\left\vert \vec{p}\right\vert
\right)  . \label{noncov}%
\end{equation}
For a spinning particle we include spin degrees of freedom in an enlarged
phase space. Of course, a massless particle is a relativistically invariant
system, and this is verified by the fact that this action is invariant under
Lorentz transformations SO$\left(  d-1,1\right)  $. However, this symmetry is
only partially manifest in this action: rotation symmetry SO$\left(
d-1\right)  $ is evident while the boost symmetry is hidden. To make the
Lorentz symmetry fully manifest one must introduce a gauge symmetry together
with extra matter gauge degrees of freedom and a gauge field. This process is
a first step in the bottom$\rightarrow$up approach that helps us discover a
deeper point of view of symmetries and their connection to spacetime.

The well known bottom-up approach in this example is to introduce the
worldline reparametrization gauge symmetry and then use the larger phase space
$\left(  x^{\mu}\left(  \tau\right)  ,p^{\mu}\left(  \tau\right)  \right)  $
with $\mu=0,1,2,\cdots,\left(  d-1\right)  .$ The action in the first order
formalism is
\begin{equation}
S\left(  x,p\right)  =\int d\tau\left(  \partial_{\tau}x^{\mu}p_{\mu}-\frac
{1}{2}ep_{\mu}p_{\nu}\eta^{\mu\nu}\right)  , \label{cov}%
\end{equation}
where $\eta_{\mu\nu}$ is the Minkowski metric and $e\left(  \tau\right)  $ is
the gauge field coupled to the generator of gauge transformations $p^{2}/2 $.
The gauge transformations
\begin{equation}
\delta_{\varepsilon}e=\partial_{\tau}\varepsilon\left(  \tau\right)
,\;\;\delta_{\varepsilon}x^{\mu}=\varepsilon\left(  \tau\right)  p^{\mu
},\;\;\delta_{\varepsilon}p_{\mu}=0,
\end{equation}
transform the Lagrangian into an ignorable total derivative $\delta
_{\varepsilon}S=\int d\tau\partial_{\tau}\left(  \varepsilon p^{2}/2\right)
=0.$ The action (\ref{cov}) has an evident global Lorentz symmetry due to the
fact that all terms are Lorentz dot products. Noether's theorem gives the
conserved charges $L^{\mu\nu}=x^{\mu}p^{\nu}-x^{\nu}p^{\mu}.$ These are gauge
invariant $\delta_{\varepsilon}L^{\mu\nu}=0$, and hence they are physical
observables. So the $L^{\mu\nu}$ remain conserved, and act as the generators
of symmetry of the gauge invariant action $S\left(  x,p\right)  ,$ even if
some arbitrary gauge is fixed.

The relation between Eqs.(\ref{noncov},\ref{cov}) is obtained in a fixed
gauge. This is the reverse process, namely it is part of the top$\rightarrow
$bottom approach that will be discussed in the next section. The equation of
motion with respect to the gauge field $e$ requires that the gauge generator
vanishes $p^{2}$=0, implying that the physical sector (massless particle on
mass shell) must be gauge invariant. The gauge symmetry can be fixed by taking
$x^{0}\left(  \tau\right)  =t\left(  \tau\right)  =\tau,$ and the constraint
can be solved for the canonical conjugate to $x^{0}=\tau,$ namely $p^{0}%
=\pm\left\vert \vec{p}\right\vert .$ The remaining phase space $\left(
\vec{x},\vec{p}\right)  $ provides a parametrization of the gauge invariant
sector. Taking the positive root $p^{0}=+\left\vert \vec{p}\right\vert $, we
derive the non-covariant action Eq.(\ref{noncov}) for the massless particle as
the gauge fixed form of the gauge invariant Eq.(\ref{cov}). Similarly, the
gauge fixed form of the gauge invariant $L^{\mu\nu},$ given by $L^{ij}%
=x^{i}p^{j}-x^{j}p^{i},$ $L^{0i}=\tau p^{i}-x^{i}\left\vert \vec{p}\right\vert
,$ are the generators of the non-linearly realized hidden global SO$\left(
d-1,1\right)  $ symmetry of the gauge fixed Lagrangian in Eq.(\ref{noncov}).
These generate the Lorentz transformations of phase space $\left(  \vec
{x}\left(  \tau\right)  ,\vec{p}\left(  \tau\right)  \right)  $ at any $\tau$
through the Poisson brackets $\delta_{\omega}\vec{x}\left(  \tau\right)
=\frac{1}{2}\omega_{\mu\nu}\left\{  L^{\mu\nu},\vec{x}\right\}  \left(
\tau\right)  ,$ and $\delta_{\omega}\vec{p}\left(  \tau\right)  =\frac{1}%
{2}\omega_{\mu\nu}\left\{  L^{\mu\nu},\vec{p}\right\}  \left(  \tau\right)  $
where $\tau$ is treated like a parameter. It can be checked that under these
transformations the non-covariant looking action in Eq.(\ref{noncov}) is
Lorentz invariant (dropping an ignorable total derivative $\delta S=\int
d\tau~\partial_{\tau}\left(  f\left(  \tau\right)  \right)  =0$).

The covariant formulation in Eq.(\ref{cov}) provides a greater flexibility to
analyze the system from a broader and more fundamental perspective. For
example, one may choose other gauges besides the timelike gauge $x^{0}\left(
\tau\right)  =\tau$ that relates Eqs.(\ref{noncov}) and (\ref{cov}). In
particular the lightcone gauge $x^{+}\left(  \tau\right)  =\tau$, in which the
constraint $p^{2}=0$ is solved for the canonical conjugate $p^{-}=p_{\perp
}^{2}/2p^{+},$ has certain advantages in computation.

One may also analyze the system covariantly. For example in covariant
quantization one may apply the constraints on the physical states to derive
the Klein-Gordon equation, and from it the Klein-Gordon free field theory%
\begin{align}
p^{2}|\varphi\rangle &  =0~\text{gauge invariant }\Leftrightarrow\text{
physical states.}\\
\langle x|p^{2}|\varphi\rangle &  =0=-\partial^{\mu}\partial_{\mu}%
\varphi\left(  x\right)  \;\rightarrow S_{KG}=\int d^{4}x~\partial^{\mu
}\varphi^{\ast}\partial_{\mu}\varphi. \label{kg}%
\end{align}
The covariant formulation in Eq.(\ref{cov}) is one of the stations in the
bottom$\rightarrow$up approach toward a deeper point of view of symmetries and spacetime.

We are not done yet with the hidden symmetries of the non-covariant action in
Eq.(\ref{noncov}). This system has the larger symmetry SO$\left(  d,2\right)
,$ namely conformal symmetry which is a general symmetry of massless systems.
This symmetry persists as a hidden symmetry in the covariant action in
Eq.(\ref{cov}) and in the Klein-Gordon action in Eq.(\ref{kg}). It has been
known that the SO$\left(  d,2\right)  $ symmetry can be made manifest in two
ways: one is twistors \cite{penrose}\cite{penrose2} (in $d=4$), and the other
is two time physics (2T-physics) \cite{2treviews}\cite{2tHandAdS} in any $d$.
Actually these are related to each other by gauge transformations in the
2T-physics formalism \cite{2ttwistor}\cite{twistorBP1} in the twistor gauge as
we will discuss in the rest of the paper in more detail. Either way, the route
to making the SO$\left(  d,2\right)  $ symmetry manifest involves introducing
a gauge symmetry, extra matter gauge degrees of freedom, and gauge fields.

\subsection{Twistor space in d=4\label{twistd4}}

The twistor formalism in $d=4$ \cite{penrose}\cite{penrose2} starts from a
different description of the massless particle. Instead of phase space degrees
of freedom $x^{\mu},p^{\mu}$ that are SO$\left(  3,1\right)  $ vectors, it
introduces SO$\left(  3,1\right)  $=SL$\left(  2,C\right)  $ spinor degrees of
freedom $Z_{A}=\left(
%TCIMACRO{\QATOP{\mu^{\dot{\alpha}}}{\lambda_{\alpha}}}%
%BeginExpansion
\genfrac{}{}{0pt}{}{\mu^{\dot{\alpha}}}{\lambda_{\alpha}}%
%EndExpansion
\right)  ,$ $A=1,2,3,4,$ constructed from SL$\left(  2,C\right)  $ doublet
spinors $\mu^{\dot{\alpha}}$, $\lambda_{\alpha},$ each described by two
complex degrees of freedom $\alpha,\dot{\alpha}=1,2.$ The quartet $Z_{A}$ is
the spinor representation $\mathbf{4}$ of the conformal group SO$\left(
4,2\right)  =$SU$\left(  2,2\right)  $, while its conjugate $\bar{Z}%
^{A}=\left(  Z^{\dagger}C\right)  ^{A}=\left(  \bar{\lambda}_{\dot{\alpha}%
}~\bar{\mu}^{\alpha}\right)  ,$with the SU$\left(  2,2\right)  $ metric
\begin{equation}
C=\sigma_{1}\times1, \label{C}%
\end{equation}
is the anti-quartet $\mathbf{\bar{4}}$ that corresponds to the second spinor
representation of SO$\left(  4,2\right)  .$ An over-bar such as $\bar{\lambda
}_{\dot{\alpha}}$ means complex conjugate of $\lambda_{\alpha}.$ The spinor is
subject to a SU$\left(  2,2\right)  $ invariant helicity constraint $Z_{A}%
\bar{Z}^{A}=\mu^{\dot{\alpha}}\bar{\lambda}_{\dot{\alpha}}+\lambda_{\alpha
}\bar{\mu}^{\alpha}=2h,$ where $h$ is the helicity of the particle. The
helicity constraint is the generator of a U$\left(  1\right)  $ gauge symmetry
that acts on the twistor through the local phase transformation $Z_{A}\left(
\tau\right)  \rightarrow Z_{A}^{\prime}\left(  \tau\right)  =e^{i\Lambda
\left(  \tau\right)  }Z_{A}\left(  \tau\right)  .$ The gauge invariant action
that describes the dynamics of twistors in four dimensions is%
\begin{equation}
S\left(  Z\right)  =\int d\tau\left(  i\bar{Z}^{A}DZ_{A}-2hV\right)
,\;\;DZ_{A}\equiv\frac{\partial Z_{A}}{\partial\tau}-iVZ_{A}. \label{action}%
\end{equation}
Here the 1-form $Vd\tau$ is a U$\left(  1\right)  $ gauge field on the
worldline, $DZ_{A}$ is the gauge covariant derivative that satisfies
$\delta_{\Lambda}\left(  DZ_{A}\right)  =i\Lambda\left(  \tau\right)  \left(
DZ_{A}\right)  $ for $\delta_{\Lambda}V=\partial\Lambda/\partial\tau$ and
$\delta_{\Lambda}Z_{A}=i\Lambda\left(  \tau\right)  Z_{A}$. Note that the term
$2hV$ (absent in previous literature) is gauge invariant since it transforms
as a total derivative under the gauge transformation$.$ The reason for
requiring the U$\left(  1\right)  $ gauge symmetry is the fact that the
overall phase of the $Z_{A}$ is unphysical and drops out in the relation
between phase space and twistors, as in Eq.(\ref{penrose}). Furthermore, the
equation of motion with respect to $V$ imposes the constraint $Z_{A}\bar
{Z}^{A}-2h=0,$ which is interpreted as the helicity constraint. Taking into
account that $\left(  Z_{A}\bar{Z}^{A}-2h\right)  $ is the generator of the
U$\left(  1\right)  $ gauge transformations, the meaning of the vanishing
generator (or helicity constraint) is that only the U$\left(  1\right)  $
gauge invariant sector of twistor space is physical.

To establish the equivalence between the massless spinless particle in
Eqs.(\ref{noncov},\ref{cov}) in $d=4,$ and twistors with vanishing helicity
$h=0,$ we must choose a U$\left(  1\right)  $ gauge for $Z_{A}$ and solve the
constraint $Z_{A}\bar{Z}^{A}=\mu^{\dot{\alpha}}\bar{\lambda}_{\dot{\alpha}%
}+\lambda_{\alpha}\bar{\mu}^{\alpha}=0.$ First we count degrees of freedom.
The twistor $Z_{A}$ has 4 complex , or 8 real, degrees of freedom. Gauge
fixing the U$\left(  1\right)  $ symmetry and solving the helicity constraint
removes 2 real degrees of freedom, leaving behind 6 real degrees of freedom,
which is the same number as the phase space degrees of freedom $\left(
\vec{x},\vec{p}\right)  .$

More explicitly, Penrose has provided the transformation between twistors and
the phase space of spinless massless particles as follows%
\begin{equation}
\mu^{\dot{\alpha}}=-ix^{\dot{\alpha}\beta}\lambda_{\beta},\;\lambda_{\alpha
}\bar{\lambda}_{\dot{\beta}}=p_{\alpha\dot{\beta}}, \label{penrose}%
\end{equation}
where the $2\times2$ Hermitian matrices $x^{\dot{\alpha}\beta},$
$p_{\alpha\dot{\beta}}$ are expanded in terms of the Pauli matrices%
\begin{equation}
x^{\dot{\alpha}\beta}\equiv\frac{1}{\sqrt{2}}x^{\mu}\left(  \bar{\sigma}_{\mu
}\right)  ^{\dot{\alpha}\beta},\;p_{\alpha\dot{\beta}}\equiv\frac{1}{\sqrt{2}%
}p^{\mu}\left(  \sigma_{\mu}\right)  _{\alpha\dot{\beta}};\;\sigma_{\mu}%
\equiv\left(  1,\vec{\sigma}\right)  ,\;\bar{\sigma}_{\mu}\equiv\left(
-1,\vec{\sigma}\right)  .
\end{equation}
Here $\lambda_{\alpha}$ can be gauge fixed by choosing a phase, and the
helicity constraint is explicitly solved since
\begin{equation}
\bar{Z}^{A}Z_{A}=\left(  \bar{\lambda}_{\dot{\alpha}}~\bar{\mu}^{\alpha
}\right)  \left(
%TCIMACRO{\QATOP{\mu^{\dot{\alpha}}}{\lambda_{\alpha}}}%
%BeginExpansion
\genfrac{}{}{0pt}{}{\mu^{\dot{\alpha}}}{\lambda_{\alpha}}%
%EndExpansion
\right)  =\bar{\lambda}_{\dot{\alpha}}\mu^{\dot{\alpha}}+\bar{\mu}^{\alpha
}\lambda_{\alpha}=-i\bar{\lambda}_{\dot{\alpha}}x^{\dot{\alpha}\beta}%
\lambda_{\beta}+i\bar{\lambda}_{\dot{\beta}}x^{\dot{\beta}\alpha}%
\lambda_{\alpha}=0. \label{tw1}%
\end{equation}
Furthermore, since $\lambda_{\alpha}\bar{\lambda}_{\dot{\beta}}$ is a
2$\times2$ matrix of rank one, its determinant vanishes. Then the
parametrization $p_{\alpha\dot{\beta}}=\lambda_{\alpha}\bar{\lambda}%
_{\dot{\beta}}$ of the momentum $p^{\mu},$ with $\sqrt{2}p^{0}=Tr\left(
p\right)  =\bar{\lambda}\lambda,$ insures automatically that $p^{0}>0$ and
that the mass shell condition is satisfied $\det\left(  \lambda\bar{\lambda
}\right)  =\det\left(  p\right)  =p^{\mu}p_{\mu}=0.$ Furthermore, by inserting
the twistor transform in Eq.(\ref{penrose}) into the twistor action $S\left(
Z\right)  $ we recover the massless particle action Eq.(\ref{cov}) on which
$p^{2}=0$ is already imposed by the form of $p^{\mu}$
\begin{align}
S\left(  Z\right)   &  =\int d\tau\left[  \bar{\lambda}_{\dot{\alpha}}%
\frac{\partial}{\partial\tau}\left(  x^{\dot{\alpha}\beta}\lambda_{\beta
}\right)  -\bar{\lambda}_{\dot{\alpha}}x^{\dot{\alpha}\beta}\frac
{\partial\lambda_{\beta}}{\partial\tau}+0\right]  =\int d\tau\frac{\partial
x^{\dot{\alpha}\beta}}{\partial\tau}\lambda_{\beta}\bar{\lambda}_{\dot{\alpha
}}\\
&  =\frac{1}{2}\int d\tau\frac{\partial x_{\mu}}{\partial\tau}p_{\nu}Tr\left(
\bar{\sigma}^{\mu}\sigma^{\nu}\right)  =\int d\tau\frac{\partial x_{\mu}%
}{\partial\tau}p_{\mu},\;\text{(with }p^{2}=0\text{)}%
\end{align}
Finally, fixing the $\tau$ reparametrization symmetry $x^{0}=\tau$ shows that
$S\left(  Z\right)  $ gives the original massless particle action
Eq.(\ref{noncov}). From this we conclude that the canonical structure of
twistors determined by $S\left(  Z\right)  ,$ namely $\left[  Z_{A},\bar
{Z}^{B}\right]  =\delta_{A}^{~B},$ is equivalent to the canonical structure in
phase space determined by $S\left(  x,p\right)  ,$ namely $\left[  x^{\mu
},p_{\nu}\right]  =i\delta_{~\nu}^{\mu},$ in the gauge invariant sectors
determined respectively by the constraints $\bar{Z}Z=0$ and $p^{2}=0$.

The analog of covariant quantization described in Eq.(\ref{kg}) can also be
done in twistor space. At the quantum level the Hermitian ordered product is
applied on physical states $\frac{1}{2}(Z_{A}\bar{Z}^{A}+\bar{Z}^{A}%
Z_{A})|\psi\rangle=2h|\psi\rangle$. Wavefunctions in twistor space are
obtained as $\psi\left(  Z\right)  =\langle Z|\psi\rangle$ where the operators
$Z_{A}$ are diagonalized on the states labelled as $\langle Z|$. Using the
fact that the canonical conjugate acts as $\bar{Z}^{A}\psi\left(  Z\right)
=\langle Z|\bar{Z}^{A}|\psi\rangle=-\frac{\partial}{\partial Z_{A}}\psi\left(
Z\right)  ,$ the physical state condition in $Z$ space $\frac{1}{2}\langle
Z|(Z_{A}\bar{Z}^{A}+\bar{Z}^{A}Z_{A})|\psi\rangle=2h\langle Z|\psi\rangle$
produces Penrose's homogeneity constraint, $Z_{A}\frac{\partial}{\partial
Z_{A}}\psi\left(  Z\right)  =\left(  -2h-2\right)  \psi\left(  Z\right)  ,$
known to correctly describe the quantum wavefunction $\psi\left(  Z\right)  $
of a particle of helicity $h.$

As the analog of the Klein-Gordon field theory of Eq.(\ref{kg}), we propose an
action in twistor space field theory that takes into account at once both the
positive and negative helicities
\begin{equation}
S_{h}\left(  \psi\right)  =\int d^{4}Z~\psi^{\ast}\left(  Z_{A}\frac{\partial
}{\partial Z_{A}}+2h+2\right)  \psi. \label{Sh}%
\end{equation}
The minimal action principle yields the homogeneity constraints. Indeed, the
equations of motion derived from $S_{h}\left(  \psi\right)  $ for $\psi
,\psi^{\ast}$ show that $\psi\left(  Z\right)  $ is the helicity $+h$
wavefunction, while $\psi^{\ast}\left(  Z\right)  $ is the helicity $-h$
wavefunction, so that together they describe a CPT invariant free field
theory. Therefore $S_{h}\left(  \psi\right)  $ is the twistor equivalent of
the Klein-Gordon, Dirac, Maxwell and higher spinning particle free field
actions in four flat dimensions. The free field theory $S_{h}\left(
\psi\right)  $ is evidently invariant under conformal transformations
SU$\left(  2,2\right)  $ by taking $\psi\left(  Z\right)  $ to transform like
a scalar while $Z_{A}$ transforms like the fundamental representation of
SU$\left(  2,2\right)  .$ Our field theory proposal for an action principle
$S_{h}\left(  \psi\right)  $ for any spinning particle seems to be new in the literature.

What do we learn from the twistor approach? Perhaps, as Penrose would suggest,
twistors may be more basic than spacetime? Without a conclusive answer to that
question so far, it is nevertheless evident through the twistor program
\cite{penrose},\cite{penrose2}, and the recent twistor superstring
\cite{witten}-\cite{2tstringtwistors} that led to computational advances in
Super Yang Mills theory \cite{witten2}\cite{cachazo}, that twistor space is a
useful space, as an alternative to space-time, to discuss the physics of
massless systems.

But there is more to be said about twistors and spacetime. In addition to what
has traditionally been known about twistors, it has recently been shown
\cite{twistorBP1} that the \textit{same} twistor $Z_{A}$ in Eq.(\ref{action})
which is known to describe the on-shell massless particle, also describes a
variety of other on-shell dynamical systems. In particular the twistor
transform of Eq.(\ref{penrose}) has been generalized so that the \textit{same}
twistor also gives the d=4 particle worldline actions for the massive
relativistic particle, the particle on AdS$_{4}$ or AdS$_{3}\times$S$^{1}$ or
AdS$_{2}\times$S$^{2},$ the particle on $R\times S^{3},$ the nonrelativistic
free particle in 3 space dimensions, the nonrelativistic hydrogen atom in 3
space dimensions, and a related family of other particle systems. For example
the twistor transform for the massive relativistic particle is
\cite{twistorBP1}%
\begin{equation}
\mu^{\dot{\alpha}}=-ix^{\dot{\alpha}\beta}\lambda_{\beta}\frac{2a}%
{1+a},\;\;\lambda_{\alpha}\bar{\lambda}_{\dot{\beta}}=\frac{1+a\mathbf{\;}%
}{2a\mathbf{\;}}p_{\alpha\dot{\beta}}+\frac{m^{2}}{2(x\cdot p)a}x_{\alpha
\dot{\beta}}, \label{twistormassive}%
\end{equation}
where $a\equiv\sqrt{1+\frac{m^{2}x^{2}}{\left(  x\cdot p\right)  ^{2}}}$.
Using this transform instead of Eq.(\ref{penrose}) we find that the action
$S\left(  Z\right)  $ in Eq.(\ref{action}) reduces to the action for the
massive particle $S=\int dt\left(  \partial_{t}\vec{x}\cdot\vec{p}-\sqrt
{\vec{p}^{2}+m^{2}}\right)  $ instead of the massless particle of
Eq.(\ref{noncov}). The mass parameter emerges as a modulus in relating the
twistor components $\left(  \mu^{\dot{\alpha}},\lambda_{\alpha}\right)  $ to
phase space $\left(  x^{\mu},p^{\mu}\right)  $ in a different way than
Eq.(\ref{penrose}). As seen in \cite{twistorBP1} the mass parameter can also
be thought of as the value of an extra momentum component in $4+2$ dimensions.
Similarly in the case of the twistor transform for the H-atom, a combination
of mass and the Coulomb coupling constant is a modulus, and so on for other
moduli in more general cases. We see that certain mass parameters, certain
curvature or other spacetime metric parameters, and certain coupling constants
emerge as moduli in the generalized twistor transform.

The results in \cite{twistorBP1} imply that twistor space is a unifying space
for various dynamics, with different Hamiltonians, which must be related to
one another through a web of dualities. This raises deeper questions on the
meaning of space and time, and accentuates the feeling that twistor space may
be even more fundamental than was thought of before: namely, from the point of
view of twistor space, spacetime and dynamics are emergent concepts, as
explicitly shown in the examples in ref.\cite{twistorBP1}. There seems to be a
deeper meaning for twistors in the context of unification that goes beyond the
originally envisaged role for twistors\footnote{Promoting these results to the
quantum level we must expect another remarkable result that the wave equations
that follow from our proposed action $S_{h}\left(  \psi\right)  $ in
Eq.(\ref{Sh}) must also correctly describe all the other cases unified by the
same twistor as given in \cite{twistorBP1}. This point will not be further
discussed in this paper and will be taken up in a future publication.}.

Actually ref.\cite{twistorBP1} provides a connection between the more general
twistor properties just outlined and the concept of 2T-physics. This relation
will be explained through the top$\rightarrow$down approach in the next
section. Here we will briefly describe the relevant properties of 2T-physics
that unify various particle dynamics in 1T-physics, and thus promote the
notion of spacetime to a higher level.

\subsection{2T-physics}

2T-physics can be viewed as a unification approach for one-time physics
(1T-physics) systems through higher dimensions. It is distinctly different
than Kaluza-Klein theory because there are no Kaluza-Klein towers of states,
but instead there is a family of 1T systems with duality type relationships
among them.

A particle interacting with various backgrounds in $\left(  d-1\right)  +1$
dimensions (e.g. electromagnetism, gravity, high spin fields, any potential,
etc.), usually described in a worldline formalism in 1T-physics, can be
equivalently described in 2T-physics. The 2T theory is in $d+2$ dimensions,
but has enough gauge symmetry to compensate for the extra $1+1$ dimensions, so
that the physical (gauge invariant) degrees of freedom are equivalent to those
encountered in 1T-physics.

One of the strikingly surprising aspects of 2T-physics is that a given $d+2$
dimensional 2T theory descends, through gauge fixing, down to a family of
holographic 1T images in $\left(  d-1\right)  +1$ dimensions. Each image fully
captures the gauge invariant physical content of a unique parent 2T theory,
but from the point of view of 1T-physics each image appears as a different
1T-dynamical system. The members of such a family naturally must obey
duality-type relationships among them and share many common properties. In
particular they share the same overall global symmetry in $d+2$ dimensions
that becomes hidden and non-linear when acting on the fewer $\left(
d-1\right)  +1$ dimensions in 1T-physics. Thus 2T-physics unifies many 1T
systems into a family that corresponds to a given 2T-physics parent in $d+2$ dimensions.

The essential ingredient in 2T-physics is the basic gauge symmetry Sp(2,R)
acting on phase space $X^{M},P_{M}$ in $d+2$ dimensions. The two timelike
directions is not an input, but is one of the outputs of the Sp$\left(
2,R\right)  $ gauge symmetry. A consequence of this gauge symmetry is that
position and momentum become indistinguishable at any instant, so the symmetry
is of fundamental significance. The transformation of $X^{M},P_{M}$ is
generally a nonlinear map that can be explicitly given in the presence of
background fields \cite{2tbacgrounds}, but in the absence of backgrounds the
transformation reduces to a linear doublet action of Sp$\left(  2,R\right)  $
on $\left(  X^{M},P^{M}\right)  $ for each $M$ \cite{2treviews}. The physical
phase space is the subspace that is gauge invariant under Sp$\left(
2,R\right)  .$ Since Sp$\left(  2,R\right)  $ has 3 generators, to reach the
physical space we must choose 3 gauges and solve 3 constraints. So, the gauge
invariant subspace of $d+2$ dimensional phase space $X^{M},P_{M}$ is a phase
space with six fewer degrees of freedom in $\left(  d-1\right)  $
\textit{space} dimensions $\left(  x^{i},p_{i}\right)  ,$ $i=1,2,\cdots\left(
d-1\right)  .$

In some cases it is more convenient not to fully use the three Sp$\left(
2,R\right)  $ gauge symmetry parameters and work with an intermediate space in
$\left(  d-1\right)  +1$ dimensions $\left(  x^{\mu},p_{\mu}\right)  ,$ that
includes time. This space can be further reduced to $d-1$ space dimensions
$\left(  x^{i},p_{i}\right)  $ by a remaining one-parameter gauge symmetry.

There are many possible ways to embed the $\left(  d-1\right)  +1$ or $\left(
d-1\right)  $ phase space in $d+2$ phase space, and this is done by making
Sp(2,R) gauge choices. In the resulting gauge fixed 1T system, time,
Hamiltonian, and in general curved spacetime, are emergent concepts. The
Hamiltonian, and therefore the dynamics as tracked by the emergent time, may
look quite different in one gauge versus another gauge in terms of the
remaining gauge fixed degrees of freedom. In this way, a unique 2T-physics
action gives rise to many 1T-physics systems.

The general 2T theory for a particle moving in any background field has been
constructed \cite{2tbacgrounds}. For a spinless particle it takes the form%
\begin{equation}
S=\int d\tau~\left(  \dot{X}^{M}P_{M}-\frac{1}{2}A^{ij}Q_{ij}\left(
X,P\right)  \right)  ,
\end{equation}
where the symmetric $A^{ij}\left(  \tau\right)  $ $,$ $i,j=1,2,$ is the
Sp$\left(  2,R\right)  $ gauge field, and the three Sp$\left(  2,R\right)  $
generators $Q_{ij}\left(  X\left(  \tau\right)  ,P\left(  \tau\right)
\right)  , $ which generally depend on background fields that are functions of
$\left(  X\left(  \tau\right)  ,P\left(  \tau\right)  \right)  $, are required
to form an Sp$\left(  2,R\right)  $ algebra. The background fields must
satisfy certain conditions to comply with the Sp$\left(  2,R\right)  $
requirement. An infinite number of solutions to the requirement can be
constructed \cite{2tbacgrounds}. So any 1T particle worldline theory, with any
backgrounds, can be obtained as a gauge fixed version of some 2T particle
worldline theory.

The 1T systems discussed in \cite{twistorBP1}, and alluded to in connection
with twistors above, are obtained by considering the simplest version of
2T-physics without any background fields. The 2T action for a
\textquotedblleft free\textquotedblright\ 2T particle is \cite{2treviews}
\begin{equation}
S_{2T}\left(  X,P\right)  =\frac{1}{2}\int d\tau~D_{\tau}X_{i}^{M}X_{j}%
^{N}\eta_{MN}\varepsilon^{ij}=\int d\tau~\left(  \dot{X}^{M}P^{N}-\frac{1}%
{2}A^{ij}X_{i}^{M}X_{j}^{N}\right)  \eta_{MN}. \label{2Taction}%
\end{equation}
Here $X_{i}^{M}=\left(  X^{M}~P^{M}\right)  ,$ $i=1,2,$ is a doublet under
Sp$\left(  2,R\right)  $ for every $M,$ the structure $D_{\tau}X_{i}%
^{M}=\partial_{\tau}X_{i}^{M}-A_{i}^{~j}X_{j}^{M}$ is the Sp(2,R) gauge
covariant derivative, Sp(2,R) indices are raised and lowered with the
antisymmetric Sp$\left(  2,R\right)  $ metric $\varepsilon^{ij},$ and in the
last expression an irrelevant total derivative $-\left(  1/2\right)
\partial_{\tau}\left(  X\cdot P\right)  $ is dropped from the action. This
action describes a particle that obeys the Sp$(2,R)$ gauge symmetry, so its
momentum and position are locally indistinguishable due to the gauge symmetry.
The $\left(  X^{M},P^{M}\right)  $ satisfy the Sp$\left(  2,R\right)  $
constraints
\begin{equation}
Q_{ij}=X_{i}\cdot X_{j}=0:\;X\cdot X=P\cdot P=X\cdot P=0,
\label{2Tconstraints}%
\end{equation}
that follow from the equations of motion for $A^{ij}$. The vanishing of the
gauge symmetry generators $Q_{ij}=0$ implies that the physical phase space is
the subspace that is Sp$\left(  2,R\right)  $ gauge invariant. These
constraints have non-trivial solutions only if the metric $\eta_{MN}$ has two
timelike dimensions. So when position and momentum are locally
indistinguishable, to have a non-trivial system, two timelike dimensions are
necessary as a consequence of the Sp$\left(  2,R\right)  $ gauge symmetry.

Thus the $\left(  X^{M},P^{M}\right)  $ in Eq.(\ref{2Taction}) are SO$\left(
d,2\right)  $ vectors, labelled by $M=0^{\prime},1^{\prime},\mu$ or
$M=\pm^{\prime},\mu,$ and $\mu=0,1,\cdots,\left(  d-1\right)  $ or $\mu
=\pm,1,\cdots,\left(  d-2\right)  ,$ with lightcone type definitions of
$X^{\pm^{\prime}}=\frac{1}{\sqrt{2}}\left(  X^{0^{\prime}}\pm X^{1^{\prime}%
}\right)  $ and $X^{\pm}=\frac{1}{\sqrt{2}}\left(  X^{0}\pm X^{3}\right)  .$
The SO$\left(  d,2\right)  $ metric $\eta^{MN}$ is given by
\begin{align}
ds^{2}  &  =dX^{M}dX^{N}\eta_{MN}=-2dX^{+^{\prime}}dX^{-^{\prime}}+dX^{\mu
}dX^{\nu}\eta_{\mu\nu}\\
&  =-\left(  dX^{0^{\prime}}\right)  ^{2}+\left(  dX^{1^{\prime}}\right)
^{2}-\left(  dX^{0}\right)  ^{2}+\left(  dX^{1}\right)  ^{2}+\left(
dX_{\perp}\right)  ^{2}\\
&  =-2dX^{+^{\prime}}dX^{-^{\prime}}-2dX^{+}dX^{-}+\left(  dX_{\perp}\right)
^{2}.
\end{align}
where the notation $X_{\perp}$ indicates SO$\left(  d-2\right)  $ vectors. So
the target phase space $X^{M},P_{M}$ is flat in $d+2$ dimension, and hence the
system in Eq.(\ref{2Taction}) has an SO$(d,2)$ global symmetry. The conserved
generators of SO$\left(  d,2\right)  $
\begin{equation}
L^{MN}=X^{M}P^{N}-X^{N}P^{M},\;\partial_{\tau}L^{MN}=0,
\end{equation}
commute with the SO$\left(  d,2\right)  $ invariant Sp$\left(  2,R\right)  $
generators $X\cdot X$, $P\cdot P$, $X\cdot P$.

The Sp$\left(  2,R\right)  $ local symmetry can be gauge fixed by choosing
three gauges and solving three constraints, but to keep some of the subgroups
of SO$\left(  d,2\right)  $ as evident symmetries it is more convenient to
choose two gauges and solve two constraints.

The SO$\left(  d-1,1\right)  $ covariant massless particle emerges if we
choose the two gauges, $X^{+^{\prime}}\left(  \tau\right)  =1$ and
$P^{+^{\prime}}\left(  \tau\right)  =0$, and solve the two constraints
$X^{2}=X\cdot P=0$ to obtain the $\left(  d-1\right)  +1$ dimensional phase
space $\left(  x^{\mu},p_{\mu}\right)  $ embedded in $\left(  d+2\right)  $
dimensions
\begin{align}
X^{M}  &  =\left(  \overset{+^{\prime}}{1},\;\overset{-^{\prime}}{x^{2}%
/2}~,\;\overset{\mu}{x^{\mu}}\right)  ,\label{massless1}\\
P^{M}  &  =\left(  ~0~,~x\cdot p~,\;~p^{\mu}\right)  . \label{massless2}%
\end{align}
The remaining constraint, $P^{2}=-2P^{+^{\prime}}P^{-^{\prime}}+P^{\mu}P_{\mu
}=p^{2}=0,$ which is the third Sp$\left(  2,R\right)  $ generator, remains to
be imposed on the physical sector. In this gauge the 2T-physics action in
Eq.(\ref{2Taction}) reduces to the covariant massless particle action in
Eq.(\ref{cov}). Furthermore, the Sp$\left(  2,R\right)  $ gauge invariant
$L^{MN}=X^{M}P^{N}-X^{N}P^{M}$ take the following nonlinear form
\begin{equation}
L^{\mu\nu}=x^{\mu}p^{\nu}-x^{\nu}p^{\mu},\;L^{+^{\prime}-^{\prime}}=x\cdot
p,\;L^{+^{\prime}\mu}=p^{\mu},\;L^{-^{\prime}\mu}=\frac{x^{2}}{2}p^{\mu
}-x^{\mu}x\cdot p. \label{conf0}%
\end{equation}
These are recognized as the generators of SO$\left(  d,2\right)  $ conformal
transformations of the $\left(  d-1\right)  +1$ dimensional phase space at the
classical level. Thus the conformal symmetry of the massless system is now
understood as the Lorentz symmetry in $d+2$ dimensions.

Having established the higher symmetrical version of the theory for the
massless particle as in Eq.(\ref{2Taction}) we reach a deeper level of
understanding of the symmetries as well as the presence of the $d+2$ nature of
the underlying spacetime. Furthermore we learn that the higher symmetrical
parent theory can be gauge fixed in many ways that produce not only the
massless particle system Eq.(\ref{noncov}) we started from, but also an
assortment of other particle dynamical systems, as discussed before
\cite{2treviews}\cite{2tHandAdS}\cite{twistorBP1}.

To emphasize this point we give also the massive relativistic particle gauge
by fixing two gauges and solving the constraints $X^{2}=X\cdot P=0$ explicitly
as follows%
\begin{align}
X^{M}  &  =\left(  \overset{+^{\prime}}{\frac{1+a\mathbf{\;}}{2a\mathbf{\;}}%
},\;\overset{-^{\prime}}{\;\frac{x^{2}a\mathbf{\;}}{1+a\mathbf{\;}}%
}~,~~\overset{\mu}{x^{\mu}}\right)  ,\;a\equiv\sqrt{1+\frac{m^{2}x^{2}%
}{\left(  x\cdot p\right)  ^{2}}}\label{massive2x}\\
P^{M}  &  =\left(  \frac{-m^{2}}{2(x\cdot p)a},\;\;\left(  x\cdot p\right)
a\;\mathbf{,\;\;}p^{\mu}\right)  ,\;P^{2}=p^{2}+m^{2}=0. \label{massive2p}%
\end{align}
In this gauge the 2T action reduces to the relativistic massive particle
action
\begin{equation}
S=\int d\tau~\left(  \dot{X}^{M}P^{N}-\frac{1}{2}A^{ij}X_{i}^{M}X_{j}%
^{N}\right)  \eta_{MN}=\int d\tau\left(  \dot{x}^{\mu}p_{\mu}-\frac{1}%
{2}A^{22}\left(  p^{2}+m^{2}\right)  \right)  .
\end{equation}

A little recognized fact is that this action is invariant under SO$\left(
d,2\right)  $. This SO$\left(  d,2\right)  $ does not have the form of
conformal transformations of Eq.(\ref{conf0}), but is a deformed version of
it, including the mass parameter. Its generators are obtained by inserting the
massive particle gauge into the gauge invariant $L^{MN}=X^{M}P^{N}-X^{N}P^{M}$%
\begin{align}
L^{\mu\nu}  &  =x^{\mu}p^{\nu}-x^{\nu}p^{\mu},\text{ \ \ \ }L^{+^{\prime
}-^{\prime}}=\left(  x\cdot p\right)  a,\label{LMNmassive1}\\
L^{+^{\prime}\mu}  &  =\frac{1+a\mathbf{\;}}{2a\mathbf{\;}}p^{\mu}+\frac
{m^{2}}{2\left(  x\cdot p\right)  a}x^{\mu}\label{LMNmassive2}\\
L^{-^{\prime}\mu}  &  =\frac{x^{2}a\mathbf{\;}}{1+a\mathbf{\;}}p^{\mu}-\left(
x\cdot p\right)  ax^{\mu} \label{LMNmassive3}%
\end{align}
It can be checked explicitly that the massive particle action above is
invariant under the SO$\left(  d,2\right)  $ transformations generated by the
Poisson brackets $\delta x^{\mu}=\frac{1}{2}\omega_{MN}\left\{  L^{MN},x^{\mu
}\right\}  $ and $\delta p^{\mu}=\frac{1}{2}\omega_{MN}\left\{  L^{MN},p^{\mu
}\right\}  ,$ up to a reparametrization of $A^{22}$ by a scale and an
irrelevant total derivative.

Since both the massive and massless particles give bases for the same
representation of SO$\left(  d,2\right)  $, we must expect a duality
transformation between them. Of course this transformation must be an
Sp$\left(  2,R\right)  =$SL$\left(  2,R\right)  $ local gauge transformation
$\left(
%TCIMACRO{\QATOP{\alpha}{\gamma}}%
%BeginExpansion
\genfrac{}{}{0pt}{}{\alpha}{\gamma}%
%EndExpansion%
%TCIMACRO{\QATOP{\beta}{\delta}}%
%BeginExpansion
\genfrac{}{}{0pt}{}{\beta}{\delta}%
%EndExpansion
\right)  \left(  \tau\right)  $ with unit determinant $\alpha\delta
-\beta\gamma=1,$ that transforms the doublets $\left(
%TCIMACRO{\QATOP{X^{M}}{P^{M}}}%
%BeginExpansion
\genfrac{}{}{0pt}{}{X^{M}}{P^{M}}%
%EndExpansion
\right)  \left(  \tau\right)  $ from Eqs.(\ref{massive2x},\ref{massive2p}) to
Eqs.(\ref{massless1},\ref{massless2}). The $\alpha,\beta,\gamma,\delta$ are
fixed by focussing on the doublets labelled by $M=+^{\prime}$
\begin{equation}
\left(
\begin{array}
[c]{c}%
\left(  \frac{1+a\mathbf{\;}}{2a\mathbf{\;}}\right) \\
\left(  \frac{-m^{2}}{2(x\cdot p)a}\right)
\end{array}
\right)  =\left(
\begin{array}
[c]{cc}%
\left(  \frac{1+a\mathbf{\;}}{2a\mathbf{\;}}\right)  & 0\\
\left(  \frac{-m^{2}}{2(x\cdot p)a}\right)  & \left(  \frac{2a\mathbf{\;}%
}{1+a\mathbf{\;}}\right)
\end{array}
\right)  \left(
\begin{array}
[c]{c}%
1\\
0
\end{array}
\right)  .
\end{equation}
Applying the inverse of this transformation on the doublets labelled by
$M=\mu$ gives the massless particle phase space (re-labelled by $\left(
\tilde{x}^{\mu},\tilde{p}^{\mu}\right)  $ below) in terms of the massive
particle phase space (labelled by $\left(  x^{\mu},p^{\mu}\right)  $)%
\begin{equation}
\left(
\begin{array}
[c]{cc}%
\left(  \frac{2a\mathbf{\;}}{1+a\mathbf{\;}}\right)  & 0\\
\left(  \frac{m^{2}}{2(x\cdot p)a}\right)  & \left(  \frac{1+a\mathbf{\;}%
}{2a\mathbf{\;}}\right)
\end{array}
\right)  \left(
\begin{array}
[c]{c}%
x^{\mu}\\
p^{\mu}%
\end{array}
\right)  =\left(
\begin{array}
[c]{c}%
\frac{2a\mathbf{\;}}{1+a\mathbf{\;}}x^{\mu}\\
\frac{1+a\mathbf{\;}}{2a\mathbf{\;}}p^{\mu}+\frac{m^{2}}{2(x\cdot p)a}x^{\mu}%
\end{array}
\right)  \equiv\left(
\begin{array}
[c]{c}%
\tilde{x}^{\mu}\\
\tilde{p}^{\mu}%
\end{array}
\right)
\end{equation}
This duality transformation is a canonical transformation $\left\{  \tilde
{x}^{\mu},\tilde{p}^{\nu}\right\}  =\eta^{\mu\nu}=\left\{  x^{\mu},p^{\nu
}\right\}  .$ Also note that the time coordinate $\tilde{x}^{0}$ is different
than the time coordinate $x^{0},$ and so are the corresponding Hamiltonians
for the massless particle $\tilde{p}^{0}=\sqrt{\tilde{p}^{i}\tilde{p}^{i}}$
versus the massive particle $p^{0}=\sqrt{p^{i}p^{i}+m^{2}}.$

The same reasoning applies among all gauge choices of the 2T theory in
Eq.(\ref{2Taction}). All resulting 1T dynamical systems are holographic images
of the same parent theory. The global symmetry SO$\left(  d,2\right)  $ of the
2T-physics action is shared in the same singleton\footnote{At the classical
level all Casimir eigenvalues vanish, but at the quantum level, due to
ordering of factors the Casimir eigenvalues are non-zero (see
Eq.(\ref{casimirs})) and correspond to the singleton representation.}
representation by all the emergent lower dimensional theories obtained by
different forms of gauge fixing. These include special cases of particles that
are massive or massless, relativistic and nonrelativistic, in flat or curved
spaces, free or interacting. This is an established fact in previous work on
2T-physics \cite{2treviews}\cite{2tHandAdS}, and it came into new focus by
displaying the explicit twistor$/$phase space transforms given in
\cite{twistorBP1}.

As seen above, the descendants of the $d+2$ dimensional 2T-physics action are
1T-physics dynamical systems that are dual to each other. Therefore we must
expect that they all have the same twistor representation modulo twistor gauge
transformations. This will be derived through the top$\rightarrow$down
approach in the next section.

It must be emphasized that as a by product of the top$\rightarrow$down
approach certain physical parameters, such as mass, parameters of spacetime
metric, and some coupling constants appear as moduli in the holographic image
while descending from $d+2$ dimensional phase space to $\left(  d-1\right)
+1$ dimensions or to twistors. Explicit examples of these have appeared in
\cite{twistorBP1}.

\section{Top$\rightarrow$down approach}

The 2T-physics action (\ref{2Taction}) and the twistor action (\ref{action})
in four dimensions are related to one another and can both be obtained as
gauge choices from the same theory in the 2T-physics formalism. To demonstrate
this fact and setup a general formalism for deriving the twistor transform in
any dimension, with or without supersymmetry, we discuss a unified theory that
defines the top$\rightarrow$down approach. This formalism was introduced in
\cite{2ttwistor} and developed further in the context of the twistor
superstring \cite{2tsuperstring}\cite{2tstringtwistors}. In this section we
begin without supersymmetry or compactified dimensions. These will be
introduced later.

In the case of $d=4$ the generalized twistor transform was applied explicitly
to specific cases in \cite{twistorBP1}, but the derivation of the general
formula was relegated to the present paper. In this section we will derive the
general twistor transform between twistor space $Z$ in $d$-dimensions and the
$d+2$ dimensional phase space $X^{M},P_{M}$ or $d$ dimensional phase space
$x^{\mu},p_{\mu}.$ We will show how it works explicitly in $d=3,4,5,6$ and
higher dimensions.

\subsection{SO$\left(  d,2\right)  $ local and SO$\left(  d,2\right)  $ global
symmetry\label{ggroup}}

In addition to the phase space SO$\left(  d,2\right)  $ vectors $\left(
X^{M},P^{M}\right)  ,$ we introduce a group element $g\left(  \tau\right)
\in$SO$\left(  d,2\right)  $ in the \textit{spinor} representation. It is
given by
\begin{equation}
g\left(  \tau\right)  =\exp\left(  \frac{i}{2}S^{MN}\omega_{MN}\left(
\tau\right)  \right)  =\exp\left(  \frac{1}{4}\Gamma^{MN}\omega_{MN}\left(
\tau\right)  \right)  \label{g}%
\end{equation}
The 2T particle action with Sp$\left(  2,R\right)  $ \& SO$\left(  d,2\right)
$ local and SO$\left(  d,2\right)  $ global symmetry is%
\begin{equation}
S_{2T}\left(  X,P,g\right)  =\int d\tau\left[  \frac{1}{2}\varepsilon
^{ij}\partial_{\tau}X_{i}\cdot X_{j}-\frac{1}{2}A^{ij}X_{i}\cdot X_{j}%
+\frac{4}{s_{d}}Tr\left(  ig^{-1}\partial_{\tau}gL\right)  \right]  ,\;\;
\label{S2T}%
\end{equation}
where the trace is in spinor space\footnote{The trace in spinor space gives
the dimension of the spinor $Tr\left(  1\right)  =s_{d}$ and $Tr\left(
\Gamma^{M}\bar{\Gamma}^{N}\right)  =s_{d}\eta^{MN}.$ For even dimensions
$s_{d}=2^{d/2}$ for the Weyl spinor of SO$\left(  d,2\right)  ,$ and the
$\bar{\Gamma}^{M},\Gamma^{M}$ are the gamma matrices in the bases of the two
different spinor representations$.$The correctly normalized generators of
SO$\left(  d,2\right)  $ in the spinor representation are $S^{MN}=\frac{1}%
{2i}\Gamma^{MN}$, where the even-dimension gamma matrices satisfy $\Gamma
^{M}\bar{\Gamma}^{N}+\Gamma^{N}\bar{\Gamma}^{M}=2\eta^{MN},$ while
$\Gamma^{MN}=\frac{1}{2}\left(  \Gamma^{M}\bar{\Gamma}^{N}-\Gamma^{N}%
\bar{\Gamma}^{M}\right)  $, $\Gamma^{MNK}=\frac{1}{3!}\left(  \Gamma^{M}%
\bar{\Gamma}^{N}\Gamma^{K}\mp\text{permutations}\right)  $, etc. There exists
a metric $C$ of SO$\left(  d,2\right)  $ in the spinor representation such
that when combined with hermitian conjugation it gives $C^{-1}\left(
\Gamma^{M}\right)  ^{\dagger}C=-\bar{\Gamma}^{M}$ and $C^{-1}\left(
\Gamma^{MN}\right)  ^{\dagger}C=-\Gamma^{MN}.$ So the inverse $g^{-1}$ is
obtained by combining hermitian and $C$-conjugation $g^{-1}=C^{-1}\left(
g\right)  ^{\dagger}C\equiv\bar{g}.$ In odd number of dimensions the
even-dimension gamma matrices above are combined to a larger matrix
$\hat{\Gamma}^{M}=\left(
%TCIMACRO{\QATOP{0}{\Gamma^{M}}}%
%BeginExpansion
\genfrac{}{}{0pt}{}{0}{\Gamma^{M}}%
%EndExpansion%
%TCIMACRO{\QATOP{\bar{\Gamma}^{M}}{0}}%
%BeginExpansion
\genfrac{}{}{0pt}{}{\bar{\Gamma}^{M}}{0}%
%EndExpansion
\right)  $ for $M=0^{\prime},1^{\prime},0,1,\cdots,\left(  d-2\right)  $ and
add one more matrix for the additional last dimension $\hat{\Gamma}%
^{d-1}=\left(
%TCIMACRO{\QATOP{1}{0}}%
%BeginExpansion
\genfrac{}{}{0pt}{}{1}{0}%
%EndExpansion%
%TCIMACRO{\QATOP{0}{-1}}%
%BeginExpansion
\genfrac{}{}{0pt}{}{0}{-1}%
%EndExpansion
\right)  $. The text is written as if $d$ is even; for odd dimensions we
replace everywhere $\hat{\Gamma}^{M}$ for both $\Gamma^{M}$ and $\bar{\Gamma
}^{M}.$ \label{ginverse}} and the matrix $L$ is given by%

\begin{equation}
L\equiv\frac{1}{4i}\Gamma_{MN}L^{MN}=\frac{1}{4i}\left(  \Gamma\cdot
X~\bar{\Gamma}\cdot P-\Gamma\cdot P~\bar{\Gamma}\cdot X\right)  . \label{L}%
\end{equation}
The first two terms of the action $S_{2T}\left(  X,P,g\right)  $ are the same
as Eq.(\ref{2Taction}), hence these terms are invariant under Sp$\left(
2,R\right)  $ which acts on $X_{i}^{M}=\left(  X^{M},P^{M}\right)  $ as a
doublet for every $M,$ and on $A^{ij}$ as the gauge field$.$ Furthermore, by
taking $g\left(  \tau\right)  $ as a singlet while noting that $L^{MN}%
=\varepsilon^{ij}X_{i}^{M}X_{j}^{N}=X^{M}P^{N}-X^{N}P^{M}$ is Sp$\left(
2,R\right)  $ gauge invariant$,$ we see that the full action is gauge
invariant under Sp$\left(  2,R\right)  .$ The action can be rewritten in the
form%
\begin{equation}
S_{2T}\left(  X,P,g\right)  =\int d\tau\left\{  \frac{1}{2s_{d}}%
\varepsilon^{ij}Tr\left[  \partial_{\tau}\left(  gX_{i}\cdot\Gamma
g^{-1}\right)  \left(  gX_{j}\cdot\bar{\Gamma}g^{-1}\right)  \right]
-\frac{1}{2}A^{ij}X_{i}\cdot X_{j}\right\}  .
\end{equation}
When both $X_{i}^{M}$ and $g\left(  \tau\right)  $ are transformed under local
Lorentz transformations as $\delta_{R}X_{i}^{M}=\varepsilon_{R}^{MN}\left(
\tau\right)  X_{iN}$ and $\delta_{R}g=-\frac{1}{4}\left(  g\Gamma_{MN}\right)
\varepsilon_{R}^{MN}\left(  \tau\right)  ,$ the structures $\left(
gX_{j}\cdot\Gamma g^{-1}\right)  $ and $X_{i}\cdot X_{j}$ are gauge invariant
under $\delta_{R}$. Therefore the Lagrangian has a gauge symmetry with local
SO$\left(  d,2\right)  _{R}$ parameters $\varepsilon_{R}^{MN}\left(
\tau\right)  $ when $g$ is transformed on the right side. In addition, there
is a global symmetry under SO$\left(  d,2\right)  _{L}$ when $g\left(
\tau\right)  $ is transformed from the left side as $\delta_{L}g=\frac{1}%
{4}\varepsilon_{L}^{MN}\left(  \Gamma_{MN}g\right)  $, with $\tau$ independent
parameters $\varepsilon_{L}^{MN}$.

Using Noether's theorem we construct the conserved charge $J^{MN}\left(
\tau\right)  $ of the global left side symmetry SO$\left(  d,2\right)  _{L}.$
We find $J^{MN}\sim iTr\left(  \Gamma^{MN}gLg^{-1}\right)  ,$ but we prefer to
write it in spinor space in the form
\begin{equation}
J_{A}^{~B}=\left(  gLg^{-1}\right)  _{A}^{~B}=J^{MN}\left(  \tau\right)
\left(  \frac{1}{4i}\Gamma_{MN}\right)  _{A}^{~B}. \label{J}%
\end{equation}
Note that the matrix $J_{A}^{~B}$ must have the same form as the matrix
$L_{A}^{~B}$ of Eq.(\ref{L}), i.e. $J=\left(  \frac{1}{4i}\Gamma_{MN}\right)
J^{MN}\left(  \tau\right)  ,$ since $gLg^{-1}=\frac{1}{4i}\left(  g\Gamma
_{MN}g^{-1}\right)  L^{MN}$ is a Lorentz transformation of the gamma matrices
that mixes them among themselves. By using the equations of motion for
$\left(  X,P,g\right)  $ one can show that these charges are conserved
$\partial_{\tau}J_{A}^{~B}\left(  \tau\right)  =0.$

As seen from the form of $J$ in Eq.(\ref{J}), it is gauge invariant under the
local SO$\left(  d,2\right)  _{R}$ as well as the local Sp$\left(  2,R\right)
$ transformations. Therefore the SO$\left(  d,2\right)  _{L}$ charges
$J_{A}^{~B}$ are physical observables that classify the physical states under
SO$\left(  d,2\right)  _{L}$ representations. In particular the Casimir
operators of these representations are given by $C_{n}=\frac{1}{s_{d}%
}tr\left(  \left(  2J\right)  ^{n}\right)  .$ With this in mind we study the
properties of $J.$ In particular the square of the matrix $J,$ given by
$\left(  J^{2}\right)  _{A}^{~B}=\left(  gLg^{-1}gLg^{-1}\right)  _{A}%
^{~~B}=\left(  gL^{2}g^{-1}\right)  _{A}^{~~B},$ contains important
information about the physical states as we will see below. To proceed from
here we will outline the rest of the computation of $J^{2}$ at the classical
and quantum levels.

If the square of the matrix $L^{2}$ is computed at the classical level, i.e.
not caring about the orders of generators $L_{MN},$ then one finds that
$\left(  L^{2}\right)  _{A}^{~B}$ is proportional to the identity matrix
$\delta_{A}^{B},$ $\left(  L^{2}\right)  =\left(  \frac{1}{4i}\Gamma
_{MN}L^{MN}\right)  ^{2}=\frac{1}{8}L^{MN}L_{MN}~1.$ Furthermore by computing,
still at the classical level $\frac{1}{2}L^{MN}L_{MN}=X^{2}P^{2}-\left(
X\cdot P\right)  ^{2},$ and imposing the classical constraints $X^{2}%
=P^{2}=\left(  X\cdot P\right)  =0,$ one finds that $L^{2}=0$ in the space of
gauge invariants of the classical theory. Then this implies also $J^{2}=0$ in
the space of gauge invariants of the classical theory. By taking higher powers
of $J,$ we find $J^{n}=0$ for all positive integers $n\geq2.$ Therefore all
Casimir eigenvalues are zero $C_{n}=0$ for all the classical physical
configurations of phase space. This is a special non-trivial representation of
the non-compact group SO$\left(  d,2\right)  _{L}$, and all classical gauge
invariants, which are functions of $L^{MN},$ can be classified as irreducible
multiplets of SO$\left(  d,2\right)  _{L}$.

We now consider the quantum theory. All the physical (gauge invariant) states
must fall into irreducible representations of the global symmetry SO$\left(
d,2\right)  _{L}.$ In the quantum theory the $L_{MN}$ form the Lie algebra of
SO$\left(  d,2\right)  ,$ therefore if the square of the matrix $L$ is
computed at the quantum level, by taking into account the orders of the
operators $L^{MN},$ one finds%
\begin{equation}
L^{2}=\left(  \frac{1}{4i}\Gamma_{MN}L^{MN}\right)  ^{2}=-\frac{d}{2}\left(
\frac{1}{4i}\Gamma_{MN}L^{MN}\right)  +\frac{1}{8}L^{MN}L_{MN}~1. \label{L2}%
\end{equation}
In this computation we used the properties of gamma matrices%
\[
\Gamma_{MN}\Gamma_{RS}=\Gamma_{MNRS}+\left(  \eta_{NR}\Gamma_{MS}-\eta
_{MR}\Gamma_{NS}-\eta_{NS}\Gamma_{MR}+\eta_{MS}\Gamma_{NR}\right)  +\left(
\eta_{NR}\eta_{MS}-\eta_{MR}\eta_{NS}\right)  .
\]
The term $\Gamma_{MNRS}L^{MN}L^{RS}$ vanishes for $L^{MN}=X^{[M}P^{N]}$ due to
a clash between symmetry/antisymmetry. The term \textquotedblleft$\eta
_{NR}\Gamma_{MS}\cdots$\textquotedblright\ turns into a commutator, and after
using the SO$\left(  d,2\right)  $ Lie algebra for $\left[  L^{MN}%
,L^{RS}\right]  $ it produces the linear term proportional to $d/2$ in
Eq.(\ref{L2}). The term \textquotedblleft$\eta_{NR}\eta_{MS}\cdots
$\textquotedblright\ produces the last term in Eq.(\ref{L2}). Furthermore the
Casimir $\frac{1}{2}L^{MN}L_{MN}$ does not vanish at the quantum level. As
shown in \cite{2treviews}, in the Sp$\left(  2,R\right)  $ gauge invariant
physical sector of phase space one finds that it has the fixed value $\frac
{1}{2}L^{MN}L_{MN}=1-d^{2}/4$ rather than zero. Hence, in the physical sector
of the quantum theory the matrix $J_{A}^{~B}$ satisfies the following algebra%
\begin{equation}
\left(  J^{2}\right)  _{A}^{~B}=-\frac{d}{2}~J_{A}^{~B}+\frac{1}{8}\left(
1-\frac{d^{2}}{4}\right)  ~\delta_{A}^{~B},\;\text{on physical states.}
\label{singleton}%
\end{equation}
We can compute the higher powers $J^{n}$ on physical states by repeatedly
using this relation,
\begin{equation}
\left(  J^{n}\right)  _{A}^{~B}=\alpha_{n}~J_{A}^{~B}+\beta_{n}~\delta
_{A}^{~B},
\end{equation}
and then compute the Casimir eigenvalues\footnote{Note that in the literature
one may find that the definition of the cubic and higher Casimir eigenvalues
are given as a linear combination of our $C_{n}.$} $C_{n}=\frac{1}{s_{d}%
}Tr\left(  \left(  2J\right)  ^{n}\right)  =2^{n}\beta_{n}$. Evidently the
$C_{n}$ will end up having fixed values determined by the dimension $d$ of
SO$\left(  d,2\right)  _{R}.$ In particular,
\begin{equation}
C_{2}=1-\frac{d^{2}}{4},\;\;C_{3}=d\left(  1-\frac{d^{2}}{4}\right)
,\;C_{4}=\left(  1-\frac{d^{2}}{4}\right)  \left(  1+\frac{3d^{2}}{4}\right)
,\;\text{etc.} \label{casimirs}%
\end{equation}
Therefore, at the quantum level we have identified a special unitary
representation that classifies all physical states of the theory. This is the
singleton representation of SO$\left(  d,2\right)  $ for any $d$. Our approach
shows that the singleton is more fully characterized by the constraints
satisfied by the charges in Eq.(\ref{singleton}). We will see in the next
section that these constraints will be satisfied explicitly at the quantum
level by constructing $J_{A}^{~B}$ in terms of twistors.

\subsection{Twistor gauge and the general twistor transform
\label{sttransform}}

There are different ways of choosing gauges to express the theory given by
$S_{2T}\left(  X,P,g\right)  $ in terms of the physical sector. One extreme in
gauge space is to eliminate $g$ completely, while another extreme is to
eliminate $\left(  X,P\right)  $ completely. When $g$ is eliminated we obtain
the phase space description, and when $\left(  X,P\right)  $ is eliminated we
obtain the twistor description.

Since SO$\left(  d,2\right)  _{R}$ is a local symmetry that acts on $g\left(
\tau\right)  $ from the right, $g\rightarrow g^{\prime}=gg_{R},$ and $g_{R}$
has exactly the same number of degrees of freedom as $g,$ one can gauge fix
the extended classical theory by choosing the gauge $g\left(  \tau\right)  =1.
$ In that case the theory described by Eq.(\ref{L}) in terms of $\left(
X,P,g\right)  $ reduces to the theory described by only $\left(  X,P\right)  $
in Eq.(\ref{2Taction})%
\begin{equation}
S_{2T}\left(  X,P,g\right)  \overset{g=1}{=}S_{2T}\left(  X,P\right)
,\;\;J_{A}^{~B}=L_{A}^{~B}. \label{JL}%
\end{equation}
In this gauge the conserved charge becomes $J=L,$ so that the global symmetry
SO$\left(  d,2\right)  _{L}$ becomes the SO$\left(  d,2\right)  $ global
symmetry of the $\left(  X,P\right)  $ theory. This reflects the fact that to
maintain the gauge $g=1,$ a transformation of $g$ from the left must be
compensated by a transformation from the right, therefore SO$\left(
d,2\right)  _{L}$ and SO$\left(  d,2\right)  _{R}$ become identified.

In the $g=1$ gauge there still remains the Sp$\left(  2,R\right)  $ gauge
symmetry. If one fixes this gauge as in Eqs.(\ref{massless1},\ref{massless2})
then we see that the original SO$\left(  d,2\right)  _{L}$ is interpreted, in
this gauge, as the conformal symmetry of the relativistic massless particle
given in Eqs.(\ref{conf0}). But if one fixes Sp$\left(  2,R\right)  $ as in
Eqs.(\ref{massive2x}-\ref{massive2p}) then the original SO$\left(  d,2\right)
_{L}$ is interpreted as the hidden SO$\left(  d,2\right)  $ of the massive
particle given in Eqs.(\ref{LMNmassive1}-\ref{LMNmassive3}). So, the original
SO$\left(  d,2\right)  _{L}$ applied on $g$ can take on many possible physical
interpretations as the hidden symmetry of various dynamical particle phase
spaces that arise from Sp$\left(  2,R\right)  $ gauge choices. Recall that for
all cases the conserved SO$\left(  d,2\right)  $ charges are just the physical
charges $J_{A}^{~B}=\left(  g^{-1}Lg\right)  _{A}^{~B}$ whose classical and
quantum properties were already computed in a gauge invariant way in the
previous section.

To obtain the twistor description of the system we eliminate $\left(
X^{M},P^{M}\right)  $ completely and keep only $g$ as discussed in
\cite{2ttwistor}. This is done by using the Sp$\left(  2,R\right)  $ and the
SO$\left(  d,2\right)  _{R}$ local symmetries to completely fix $X^{M},P^{M}$
to the convenient form $X^{+^{\prime}}=1$ and $P^{+}=1,$ while all other
components vanish%
\begin{equation}
X^{M}=(\overset{+^{\prime}}{{1}},\overset{-^{\prime}}{{0}},\overset{+}{{0}%
},\overset{-}{{0}},\overset{i}{{0}}),\;P^{M}=(\overset{+^{\prime}}{{0}%
},\overset{-^{\prime}}{{0}},\overset{+}{{1}},\overset{-}{{0}},\overset{i}{{0}%
}),\;i=1,\cdots,\left(  d-2\right)  . \label{twistgauge1}%
\end{equation}
These $X^{M},P^{M}$ already satisfy the constraints $X^{2}=P^{2}=X\cdot P=0$.
In this gauge the only non-vanishing component of $L^{MN}$ is $L^{+^{\prime}%
+}=1$, so that
\begin{equation}
L_{fixed}=\frac{-2}{4i}\Gamma^{-^{\prime}-}L^{+^{\prime}+}=\frac{i}{2}%
\Gamma^{-^{\prime}-}\equiv\Gamma. \label{Lfixed}%
\end{equation}
Hence the physical content of the theory is now described only in terms of $g$
and the fixed matrix $\Gamma$ embedded in the Lie algebra of SO$\left(
d,2\right)  .$

The matrix $\Gamma$ has very few non-zero entries as seen by choosing a
convenient form of gamma matrices\footnote{An explicit form of SO$\left(
d,2\right)  $ gamma matrices that we find convenient in even dimensions, is
given by $\Gamma^{0}=1\times1$,\ $\Gamma^{i}=\sigma_{3}\times\gamma^{i}%
$,\ $\Gamma^{\pm^{\prime}}=-i\sqrt{2}\sigma^{\pm}\times1$ (note $\Gamma
^{0^{\prime}}=-i\sigma_{1}\times1$ and $\Gamma^{1^{\prime}}=\sigma_{2}\times
1$), \ where $\gamma^{i}$ are the SO$\left(  d-1\right)  $ gamma matrices. The
$\bar{\Gamma}^{M}$ are the same as the $\Gamma^{M}$ for $M=\pm^{\prime},i,$
but for $M=0$ we have $\bar{\Gamma}^{0}=-\Gamma^{0}=-1\times1.$ From these we
construct the traceless $\Gamma^{+^{\prime}-^{\prime}}=\left(
%TCIMACRO{\QATOP{-1}{0}}%
%BeginExpansion
\genfrac{}{}{0pt}{}{-1}{0}%
%EndExpansion%
%TCIMACRO{\QATOP{0}{1}}%
%BeginExpansion
\genfrac{}{}{0pt}{}{0}{1}%
%EndExpansion
\right)  $,\ $\Gamma^{+^{\prime}\mu}=i\sqrt{2}\left(
%TCIMACRO{\QATOP{0}{0}}%
%BeginExpansion
\genfrac{}{}{0pt}{}{0}{0}%
%EndExpansion%
%TCIMACRO{\QATOP{\bar{\gamma}^{\mu}}{0}}%
%BeginExpansion
\genfrac{}{}{0pt}{}{\bar{\gamma}^{\mu}}{0}%
%EndExpansion
\right)  $,\ $\Gamma^{-^{\prime}\mu}=-i\sqrt{2}\left(
%TCIMACRO{\QATOP{0}{\gamma^{\mu}}}%
%BeginExpansion
\genfrac{}{}{0pt}{}{0}{\gamma^{\mu}}%
%EndExpansion%
%TCIMACRO{\QATOP{0}{0}}%
%BeginExpansion
\genfrac{}{}{0pt}{}{0}{0}%
%EndExpansion
\right)  $,\ $\Gamma^{\mu\nu}=\left(
%TCIMACRO{\QATOP{\bar{\gamma}^{\mu\nu}}{0}}%
%BeginExpansion
\genfrac{}{}{0pt}{}{\bar{\gamma}^{\mu\nu}}{0}%
%EndExpansion%
%TCIMACRO{\QATOP{0}{\gamma^{\mu\nu}}}%
%BeginExpansion
\genfrac{}{}{0pt}{}{0}{\gamma^{\mu\nu}}%
%EndExpansion
\right)  $, with $\gamma_{\mu}=\left(  1,\gamma^{i}\right)  $ and $\bar
{\gamma}_{\mu}=\left(  -1,\gamma^{i}\right)  .$ Then $\frac{1}{2}\Gamma
_{MN}J^{MN}=-\Gamma^{+^{\prime}-^{\prime}}J^{+^{\prime}-^{\prime}}$+~
$\frac{1}{2}J_{\mu\nu}\Gamma^{\mu\nu}-$ $\Gamma_{~\mu}^{+^{\prime}%
}J^{-^{\prime}\mu}-$ $\Gamma_{~\mu}^{-^{\prime}}J^{+^{\prime}\mu}$ takes the
matrix form given in Eq.(\ref{GL}). We can further write $\gamma^{1}=\tau
^{1}\times1,$ $\gamma^{2}=\tau^{2}\times1$ and $\gamma^{r}=\tau^{3}\times
\rho^{r},$ where the $\rho^{r}$ are the gamma matrices for SO$\left(
d-3\right)  $. These gamma matrices are consistent with the metric
$C=\sigma_{1}\times1\times c$ of Eq.(\ref{C}), and footnote (\ref{ginverse}),
provided $c^{-1}\left(  \rho^{r}\right)  ^{\dagger}c=\rho^{r}.$ It is possible
to choose hermitian $\rho^{r}$ with $c=1$ for SO$\left(  d-3\right)  .$ If one
works in a basis with $c\neq1,$ then hermitian conjugation of of SO$\left(
d-3\right)  $ spinors (which occur e.g. in $\bar{\lambda}$ of Eq.(\ref{zbar}))
must be supplemented by multiplying with $c,$ as in $\bar{\lambda}%
\equiv\lambda^{\dagger}\left(  1\times c\right)  .$ \label{gamms}} for
SO$\left(  d,2\right)  $. Then, up to similarity transformations, $\Gamma$ can
be brought to the form\footnote{The gamma matrices $\Gamma^{M}$ of footnote
(\ref{gamms}) can be redefined differently for the left or right sides of $g$
up to similarity transformations. Thus, for the right side of $g$ we apply a
similarity transformation so that $\gamma^{1}=\tau^{3}\times1$, etc., to
obtain $\gamma^{-}=\left(  \gamma^{0}-\gamma^{1}\right)  /\sqrt{2}$ in the
form given in Eq.(\ref{g--}).}
\begin{equation}
\Gamma=\frac{i}{2}\Gamma^{-^{\prime}-}=\frac{1}{\sqrt{2}}\left(
%TCIMACRO{\QATOP{0}{\gamma^{-}}}%
%BeginExpansion
\genfrac{}{}{0pt}{}{0}{\gamma^{-}}%
%EndExpansion%
%TCIMACRO{\QATOP{0}{0}}%
%BeginExpansion
\genfrac{}{}{0pt}{}{0}{0}%
%EndExpansion
\right)  =\left(
%TCIMACRO{\QATOP{\QATOP{0}{0}\QATOP{0}{0}}{\QATOP{0}{0}\QATOP{0}{1}}}%
%BeginExpansion
\genfrac{}{}{0pt}{}{\genfrac{}{}{0pt}{}{0}{0}\genfrac{}{}{0pt}{}{0}{0}%
}{\genfrac{}{}{0pt}{}{0}{0}\genfrac{}{}{0pt}{}{0}{1}}%
%EndExpansion%
%TCIMACRO{\QATOP{\QATOP{0}{0}\QATOP{0}{0}}{\QATOP{0}{0}\QATOP{0}{0}}}%
%BeginExpansion
\genfrac{}{}{0pt}{}{\genfrac{}{}{0pt}{}{0}{0}\genfrac{}{}{0pt}{}{0}{0}%
}{\genfrac{}{}{0pt}{}{0}{0}\genfrac{}{}{0pt}{}{0}{0}}%
%EndExpansion
\right)  . \label{g--}%
\end{equation}
The identity matrix $1,$ and the $0$'s, are $\frac{s_{d}}{4}\times\frac{s_{d}%
}{4}$ square block matrices embedded in the spinor representation of
SO$\left(  d,2\right)  .$ Then the gauge invariant 2T action in Eq.(\ref{S2T}%
), and the gauge invariant SO$\left(  d,2\right)  _{L}$ charges in
Eq.(\ref{J}), take the twistor form similar to Eq.(\ref{action})
\begin{align}
S_{2T}\left(  X,P,g\right)   &  =\frac{4}{s_{d}}\int d\tau~Tr\left(
i\partial_{\tau}g\Gamma g^{-1}\right)  =\frac{4}{s_{d}}\int d\tau
~i\partial_{\tau}Z_{A}^{~a}\bar{Z}_{a}^{~A}\equiv S_{twistor},\label{Stw}\\
\left(  J\right)  _{A}^{B}  &  =\left(  g\Gamma g^{-1}\right)  _{A}%
^{B}=\left(  Z_{A}^{~a}\bar{Z}_{a}^{~B}-\frac{1}{s_{d}}tr\left(  Z\bar
{Z}\right)  \delta_{A}^{B}\right)  ,\; \label{Jtw}%
\end{align}
The $Z_{A}^{~a},\bar{Z}_{a}^{~B}$ are the twistors that already obey the
constraints $\left(  \bar{Z}Z\right)  _{a}^{~b}=0$ in Eq.(\ref{ZZconstraint})
below, so $\frac{4}{s_{d}}\int d\tau~i\partial_{\tau}Z_{A}^{~a}\bar{Z}%
_{a}^{~A}$ is the full twistor action (for an equivalent gauge invariant
action that also produces the constraints, see Eq.(\ref{stwistor})). Due to
the form of $\Gamma$ it is useful to think of $g$ as written in the form of
$\frac{s_{d}}{4}\times\frac{s_{d}}{4}$ square blocks. Then $Z_{A}^{~a}$ with
$A=1,2,\cdots,s_{d}$ and $a=1,2,\cdots,\frac{s_{d}}{4}$ emerges as the
rectangular matrix that corresponds to the last block of columns of the matrix
$g,$ and similarly $\bar{Z}_{a}^{~A}$ corresponds to the second block of rows
of $g^{-1}.$ Since $g^{-1}=C^{-1}g^{\dagger}C,$ we find that $\bar{Z}%
=c^{-1}Z^{\dagger}C,$ where $C=\sigma_{1}\times1\times c$ is given in footnote
(\ref{gamms}). Furthermore, as part of $g,g^{-1},$ the $Z_{A}^{~a},\bar{Z}%
_{a}^{~B}$ must satisfy the constraint $\bar{Z}_{a}^{~A}Z_{A}^{~b}=0$ since
the product $\bar{Z}_{a}^{~A}Z_{A}^{~b}$ contributes to an off-diagonal block
of the matrix $1$ in $g^{-1}g=1,\;$
\begin{equation}
g^{-1}g=1\;\rightarrow\bar{Z}_{a}^{~A}Z_{A}^{~b}=0. \label{ZZconstraint}%
\end{equation}
A constraint such as this one must be viewed as the generator of a gauge
symmetry that operates on the $a$ index (the columns) of the twistor
$Z_{A}^{~a}.$

Let us do some counting of degrees of freedom. To describe the particle in $d
$ dimensions we only need $2\left(  d-1\right)  $ physical degrees of freedom
corresponding to phase space $\left(  \vec{x},\vec{p}\right)  .$ This counting
applies no matter if the particle is massless or massive, relativistic or not
relativistic, in flat space or curved space, interacting or not interacting.
Our twistors are expected to apply to all these cases, so we must have the
same number of physical parameters in the twistors given above. Any extra
parameters in $Z_{A}^{~a}$ beyond $2\left(  d-1\right)  $ must be either gauge
degrees of freedom of the twistor, or there must be additional relations among
the $Z_{A}^{~a}$. The $s_{d}\times s_{d}$ matrix $g$ (with $s_{d}=2^{d/2}$ for
even $d$) is constructed from $\frac{1}{2}\left(  d+2\right)  \left(
d+1\right)  $ group parameters $\omega_{MN}$ as in Eq.(\ref{g}), but only
$2\left(  d-1\right)  $ of those parameters, corresponding to a coset,
contribute in Eqs.(\ref{Stw},\ref{Jtw}) due to the form of $\Gamma$ as will be
explained in section (\ref{coset}). So for sufficiently large $d,$ we expect
to find many relations among the $s_{d}^{2}/4=2^{2d-2}$ entries in the
rectangular matrix $Z_{A}^{~a}\left(  g\right)  .$

In $d=3,$ with $s_{3}/4=1$ columns, the $Z_{A}$ form the fundamental
representation of Sp$\left(  4,R\right)  =$SO$\left(  3,2\right)  .$ The
single $Z_{A}$ has just 4 real components that automatically satisfy the
constraint $\bar{Z}^{~A}Z_{A}=0.$ This number of degrees of freedom precisely
matches the expected number 4 of physical degrees of freedom, $2\left(
d-1\right)  =2\left(  3-1\right)  =4,$ for $d=3.$ So there are no extra
relations among the 4 real twistor entries in $Z_{A}.$

In $d=4$ dimensions, with s$_{4}/4=1$ columns, the $Z_{A}$ form the
fundamental representation of SU$\left(  2,2\right)  =$SO$\left(  4,2\right)
.$ The single $Z_{A}$ has 4 complex components or 8 real parameters that must
satisfy the U$\left(  1\right)  $ gauge constraint $\bar{Z}^{~A}Z_{A}=0$ of
Eq.(\ref{ZZconstraint}). The U$\left(  1\right)  $ gauge symmetry together
with the constraint remove 2 real parameters. So the $Z_{A}$ contains $8-2=6$
physical degrees of freedom, which is just the correct number $2\left(
4-1\right)  =6$ in $d=4,$ as discussed in section (\ref{twistd4}). So there
are no extra relations among the 4 complex twistor entries in $Z_{A}.$

For $d=5,6,$ with $s_{5}/4=s_{6}/4=2$ columns, the $Z_{A}^{~a}$ is a doublet
under an SU$\left(  2\right)  $ gauge symmetry for $d=6,$ and SU$\left(
2\right)  \times$U$\left(  1\right)  $ for $d=5$. Beyond the SU$\left(
2\right)  $ or SU$\left(  2\right)  \times$U$\left(  1\right)  $ gauge freedom
there seems to be further relations, but these amount to a simple
pseudo-reality condition on $Z_{A}^{~a},$ consistent with the transformation
rules of $Z_{A}^{~a}$ under Spin$\left(  6,2\right)  \times$SU$\left(
2\right)  $. The pseudo-reality condition emerges from the pseudo-reality of
the spinor representation of SO$\left(  6,2\right)  $ or SO$\left(
5,2\right)  . $ So, again there are no complicated relations among the entries
of $Z_{A}^{~a}$ for $d=5,6$ as seen by the following simple counting. The
number of real entries in the pseudo-real $Z_{A}^{~a}$ is $s_{6}^{2}%
/8=s_{5}^{2}/8=8^{2}/4=16.$ The 3 SU$\left(  2\right)  $ gauge conditions
together with the 3 constraints remove 6 parameters, leaving $16-6=10$, which
is precisely the correct number of physical degrees of freedom for $d=6,$ i.e.
$2\left(  6-1\right)  =10.$ Similarly, for $d=5$ the extra U$\left(  1\right)
$ and its constraint removes two more real parameters and this matches the
correct number $2\left(  5-1\right)  =8.$

For higher dimensions there are gauge symmetries among the columns but there
also are further complicated relations among the $Z_{A}^{~a}$. It turns out
that in all cases $d\geq3$ just the first column of $Z_{A}^{~a}$ (i.e. $a=1$)
already contains all the parameters that describe the physical degrees of
freedom, but it is still useful to deal with the full $Z_{A}^{~a}$ since all
components can be conveniently given explicitly in terms of gamma matrices, as
seen in Eqs.(\ref{mu},\ref{lambd}) below.

We can construct explicitly the $Z_{A}^{~a},\bar{Z}_{a}^{~A}$ that satisfy all
of the relations discussed above at the classical level. This will give the
twistor transform we are after. The key is the gauge invariant $J_{A}^{B}.$ We
identify its two different forms in the two different gauges, one in terms of
phase space and the other in terms of twistors, as given in Eqs.(\ref{JL}%
,\ref{Jtw})
\begin{equation}
\frac{1}{4i}\Gamma_{MN}L^{MN}\;\overset{g=1}{=}\;J\;\overset{X,P\sim0}%
{=}\;Z\bar{Z},\;\text{with }\left(  \bar{Z}Z\right)  _{a}^{~b}=0,~Tr\left(
Z\bar{Z}\right)  =0.\;\; \label{basicrelation}%
\end{equation}
Of course, the $\left(  X,P,g=1\right)  $ on the left side of the equation are
gauge transformations of the ($X^{M}=\delta_{+^{\prime}}^{M}$, $P^{M}%
=\frac{s_{d}}{4}\delta_{+}^{M}$, and $Z\left(  g\right)  $) on the right side.
So this equation must contain the twistor transform. More explicitly we write
$Z$ in terms of its components
\begin{equation}
Z_{A}^{~a}=\left(
\begin{array}
[c]{c}%
\mu\\
\lambda
\end{array}
\right)  ,\;\bar{Z}_{a}^{~A}=\left(
\begin{array}
[c]{cc}%
\bar{\lambda} & \bar{\mu}%
\end{array}
\right)  ,\;\bar{Z}_{a}^{~A}Z_{A}^{~b}=\left(  \bar{\lambda}\mu+\bar{\mu
}\lambda\right)  _{a}^{~b}=0, \label{zbar}%
\end{equation}
where $\mu,\lambda$ are $\frac{s_{d}}{2}\times\frac{s_{d}}{4}$ rectangular
matrices. Then we use the gamma matrices, and the definition of $\bar{\lambda
},\bar{\mu}$ in footnote (\ref{gamms}), to write the basic relation
(\ref{basicrelation}) between phase space and twistors more explicitly as%
\begin{equation}
\frac{1}{2i}\left(
\begin{array}
[c]{cc}%
L^{+^{\prime}-^{\prime}}+\frac{1}{2}L_{\mu\nu}\bar{\gamma}^{\mu\nu} &
-i\sqrt{{2}}L^{-^{\prime}\mu}~\bar{\gamma}_{\mu}\\
i\sqrt{{2}}L^{+^{\prime}\mu}~\gamma_{\mu} & -L^{+^{\prime}-^{\prime}}+\frac
{1}{2}L_{\mu\nu}\gamma^{\mu\nu}%
\end{array}
\right)  =J=\left(
\begin{array}
[c]{cc}%
\mu\bar{\lambda} & \mu\bar{\mu}\\
\lambda\bar{\lambda} & \lambda\bar{\mu}%
\end{array}
\right)  . \label{GL}%
\end{equation}
Comparing the lower off diagonal blocks we learn part of the twistor transform%
\begin{equation}
\lambda\bar{\lambda}=\frac{1}{\sqrt{2}}L^{+^{\prime}\mu}~\gamma_{\mu}=\frac
{1}{\sqrt{2}}\left(  X^{+^{\prime}}P^{\mu}-P^{+^{\prime}}X^{\mu}\right)
\gamma_{\mu}. \label{lambda}%
\end{equation}
We should also note the twistor relations that follow from the other three
blocks. In three or four dimensions a single doublet $\lambda$ satisfies this
equation automatically. In higher dimensions a single column $\lambda$ cannot
do it automatically, instead we have $\frac{s_{d}}{4}$ columns in $\lambda$
with certain relations among them. Thanks to the relations among columns, that
can be expressed in terms of gamma matrices as in Eqs.(\ref{lambd}) below, the
equation above will be satisfied.

Next we consider $\left(  \bar{\Gamma}\cdot X\right)  J$ and use the different
gauge fixed forms of the gauge invariant $J$ to show that it vanishes as
follows%
\begin{align}
\left(  \bar{\Gamma}\cdot X\right)  J  &  =\left(  \bar{\Gamma}\cdot X\right)
L=\frac{1}{4i}\left(  \bar{\Gamma}\cdot X\right)  \left(  \Gamma\cdot
X~\bar{\Gamma}\cdot P-\Gamma\cdot P~\bar{\Gamma}\cdot X\right) \nonumber\\
&  =\frac{1}{2i}X\cdot X\left(  \bar{\Gamma}\cdot P\right)  -\frac{1}%
{4i}X\cdot P~\left(  \bar{\Gamma}\cdot X\right)  =0\text{.}%
\end{align}
The last zero is because $X\cdot P=P\cdot P=0$ in the Sp$\left(  2,R\right)  $
gauge invariant physical sector. Similarly we show also $\left(  \bar{\Gamma
}\cdot P\right)  J=0.$ Therefore
\begin{equation}
\left(  \bar{\Gamma}\cdot P\right)  J=0,\;\left(  \bar{\Gamma}\cdot X\right)
J=0.
\end{equation}
Hence, every column of $J$ is a simultaneous null eigenstate of the matrices
$\left(  \bar{\Gamma}\cdot X\right)  $ and $\left(  \bar{\Gamma}\cdot
P\right)  . $

Furthermore, because $J$ can be written in the form $J=Z\bar{Z},$ it must be
that the $s_{d}\times\frac{s_{d}}{4}$ matrix $Z$ is a collection of these null
eigenstates, so it is possible to write $Z$ as a linear combination of
$\frac{s_{d}}{4}$ columns of $L=J$ as follows
\begin{equation}
Z_{A}^{~a}=\left(
%TCIMACRO{\QATOP{\mu}{\lambda}}%
%BeginExpansion
\genfrac{}{}{0pt}{}{\mu}{\lambda}%
%EndExpansion
\right)  =L\left(
%TCIMACRO{\QATOP{\QATOP{\alpha_{1}}{\alpha_{2}}}{\QATOP{\alpha_{3}}{\alpha_{4}%
%}}}%
%BeginExpansion
\genfrac{}{}{0pt}{}{\genfrac{}{}{0pt}{}{\alpha_{1}}{\alpha_{2}}%
}{\genfrac{}{}{0pt}{}{\alpha_{3}}{\alpha_{4}}}%
%EndExpansion
\right)  ,\text{ } \label{ZJa}%
\end{equation}
where $\alpha_{i}$ are $\frac{s_{d}}{4}\times\frac{s_{d}}{4}$ matrices that
will be determined below by requiring $J=L=Z\bar{Z}$. So $Z$ must also satisfy
the null conditions
\begin{equation}
\left(  \bar{\Gamma}\cdot P\right)  Z=0,\text{ }\;\left(  \bar{\Gamma}\cdot
X\right)  Z=0\;\text{and }\;LZ=0. \label{null}%
\end{equation}
The last null relation $LZ=0$ is satisfied automatically if the first two are
satisfied. This property is sufficient to determine $Z_{A}^{~a}$ up to gauge
transformations as follows. The explicit matrices are obtained by using the
gamma matrices in footnote (\ref{gamms})
\begin{equation}
\left(  \bar{\Gamma}\cdot X\right)  =\left(
\begin{array}
[c]{cc}%
X_{\mu}\gamma^{\mu} & -i\sqrt{2}X^{-^{\prime}}\\
-i\sqrt{2}X^{+^{\prime}} & -X_{\mu}\bar{\gamma}^{\mu}%
\end{array}
\right)  ,\;\;\left(  \bar{\Gamma}\cdot P\right)  =\left(
\begin{array}
[c]{cc}%
P_{\mu}\gamma^{\mu} & -i\sqrt{2}P^{-^{\prime}}\\
-i\sqrt{2}P^{+^{\prime}} & -P_{\mu}\bar{\gamma}^{\mu}%
\end{array}
\right)
\end{equation}
Then the zero eigenvalue conditions $\left(  \bar{\Gamma}\cdot X\right)
Z=0=\left(  \bar{\Gamma}\cdot P\right)  Z$ are solved by%
\begin{equation}
\mu=-i\frac{X_{\mu}\bar{\gamma}^{\mu}}{\sqrt{2}X^{+^{\prime}}}\lambda
=-i\frac{P_{\mu}\bar{\gamma}^{\mu}}{\sqrt{2}P^{+^{\prime}}}\lambda
\;,\;\;\lambda=i\frac{X_{\mu}\gamma^{\mu}}{\sqrt{2}X^{-^{\prime}}}\mu
=i\frac{P_{\mu}\gamma^{\mu}}{\sqrt{2}P^{-^{\prime}}}\mu. \label{mulambda}%
\end{equation}
To show that these expressions are consistent with each other, note that the
second set is obtained by inverting the first set as long as $X^{2}%
=P^{2}=X\cdot P=0$ are satisfied in the physical sector. For example, multiply
both sides of the equation $\mu=-i\frac{X_{\mu}\bar{\gamma}^{\mu}}{\sqrt
{2}X^{+^{\prime}}}\lambda$ by $i\frac{X_{\mu}\gamma^{\mu}}{\sqrt
{2}X^{-^{\prime}}},$ then use $\bar{\gamma}^{\mu}\gamma^{\nu}+\bar{\gamma
}^{\nu}\gamma^{\mu}=2\eta^{\mu\nu}$ and $X^{2}=X^{\mu}X_{\mu}-2X^{+^{\prime}%
}X^{-^{\prime}}=0,$ to obtain $\;\lambda=i\frac{X_{\mu}\gamma^{\mu}}{\sqrt
{2}X^{-^{\prime}}}\mu.$ So we can concentrate on the consistency of the first
set only. The difference between the two expressions for $\mu$ must vanish,
this implies that $\lambda$ should satisfy the Dirac-like equation%
\begin{equation}
\left(  X^{+^{\prime}}P^{\mu}-P^{+^{\prime}}X^{\mu}\right)  \bar{\gamma}_{\mu
}\lambda=L^{+^{\prime}\mu}~\gamma_{\mu}\lambda=0. \label{lambdaeq}%
\end{equation}
This is a consistent equation provided the vector $L^{+^{\prime}\mu}~$is null
\begin{equation}
L^{+^{\prime}\mu}L_{~\mu}^{+^{\prime}}=\left(  X^{+^{\prime}}P^{\mu
}-P^{+^{\prime}}X^{\mu}\right)  ^{2}=0.
\end{equation}
This is indeed correct in the physical sector that satisfies $X^{2}%
=P^{2}=X\cdot P=0.$ Then we can solve for $\lambda$ by writing%
\begin{equation}
\lambda=\left(  X^{+^{\prime}}P^{\mu}-P^{+^{\prime}}X^{\mu}\right)
\gamma_{\mu}\left(
%TCIMACRO{\QATOP{\alpha}{\beta}}%
%BeginExpansion
\genfrac{}{}{0pt}{}{\alpha}{\beta}%
%EndExpansion
\right)  .
\end{equation}
This satisfies the $\lambda$ equation (\ref{lambdaeq}) automatically for any
$\frac{s_{d}}{4}\times\frac{s_{d}}{4}$ matrices $\alpha,\beta$. These in turn
are determined, up to $\frac{s_{d}}{4}\times\frac{s_{d}}{4}$ gauge
transformations, by satisfying the $\frac{s_{d}}{4}\times\frac{s_{d}}{4}$
matrix equation (\ref{lambda}).

In this way we have seen that all of the forms given in Eq.(\ref{mulambda})
are consistent with each other and determine $Z_{A}^{~a}$ up to a gauge
transformation. Thus the null conditions in Eq.(\ref{null}) is all that is
needed to determine $Z_{A}^{~a}$ up to a gauge transformation, but in turn
these followed from the basic relation in Eq.(\ref{GL}).

The following subset of our relations resemble the twistor transform in four
dimensions, but $\mu,\lambda$ are $\frac{s_{d}}{2}\times\frac{s_{d}}{4}$
matrices that must obey all the relations above
\begin{equation}
\mu=-i\frac{X_{\mu}\bar{\gamma}^{\mu}}{\sqrt{2}X^{+^{\prime}}}\lambda
,\;\text{and ~}\lambda\bar{\lambda}=\frac{1}{\sqrt{2}}\left(  X^{+^{\prime}%
}P^{\mu}-P^{+^{\prime}}X^{\mu}\right)  \gamma_{\mu}. \label{tttransf}%
\end{equation}
Indeed we can check directly that by inserting only these relations into the
right hand side of Eqs.(\ref{GL},\ref{basicrelation}), we derive the
SO$\left(  d,2\right)  $ generators in terms of phase space $L^{MN}=X^{M}%
P^{N}-X^{N}P^{N}$ that appear on the left side of those equations. Furthermore
by inserting only these relations into the twistor action we derive the phase
space action that determines the canonical structure%
\begin{align}
\frac{4}{s_{d}}\int d\tau~i\partial_{\tau}Z_{A}^{~a}\bar{Z}_{a}^{~A}  &
=i\frac{4}{s_{d}}\int d\tau~Tr\left(  \partial_{\tau}\mu\bar{\lambda}%
+\partial_{\tau}\lambda\bar{\mu}\right)  =\frac{4}{s_{d}}\int d\tau~Tr\left(
\partial_{\tau}\left(  \frac{X_{\mu}\bar{\gamma}^{\mu}}{\sqrt{2}X^{+^{\prime}%
}}\right)  \lambda\bar{\lambda}\right) \\
&  =\frac{4}{s_{d}}\int d\tau~\frac{1}{\sqrt{2}}\left(  X^{+^{\prime}}P^{\mu
}-P^{+^{\prime}}X^{\mu}\right)  Tr\left(  \partial_{\tau}\left(  \frac{X_{\mu
}\bar{\gamma}^{\mu}}{\sqrt{2}X^{+^{\prime}}}\right)  \gamma_{\mu}\right) \\
&  =\int d\tau~\left(  X^{+^{\prime}}P_{\mu}-P^{+^{\prime}}X_{\mu}\right)
\partial_{\tau}\left(  \frac{X^{\mu}}{X^{+^{\prime}}}\right) \\
&  =\int d\tau~\left(  P_{\mu}-\frac{P^{+^{\prime}}}{X^{+^{\prime}}}X_{\mu
}\right)  \left(  \partial_{\tau}X^{\mu}-\frac{\partial_{\tau}X^{+^{\prime}}%
}{X^{+^{\prime}}}X^{\mu}\right) \\
&  =\int d\tau~\left(  \partial_{\tau}X^{\mu}P_{\mu}-\partial_{\tau
}X^{+^{\prime}}P^{-^{\prime}}-\partial_{\tau}X^{-^{\prime}}P^{+^{\prime}%
}\right)  =\int d\tau~\partial_{\tau}X^{M}P_{M}.
\end{align}
The last line follows when the constraints $X^{2}=P^{2}=X\cdot P=0$ are
satisfied in the physical sector. This shows the consistency of our twistor
transform of Eq.(\ref{tttransf}) for spinless particles in all dimensions.

It is also interesting to give an explicit formula for both $\mu$ and
$\lambda$ in terms of the Sp$\left(  2,R\right)  $ gauge invariant $L^{MN}.$
This is already obtained through the relation between $Z$ and $L$ given in
Eq.(\ref{ZJa}). It turns out that it is sufficient to take only one of the
$\alpha_{i}$ to be nonzero. So we will take $\alpha_{2}=\alpha_{3}=\alpha
_{4}=0$ and determine $\alpha_{1}\neq0$ from the relation $J=L=Z\bar{Z}$. The
other possibilities are gauge equivalent. The equivalence is guaranteed by the
Sp$\left(  2,R\right)  $ gauge invariance conditions $X^{2}=P^{2}=X\cdot P=0$.
This gives%
\begin{equation}
Z_{A}^{~a}=\left(
%TCIMACRO{\QATOP{\mu}{\lambda}}%
%BeginExpansion
\genfrac{}{}{0pt}{}{\mu}{\lambda}%
%EndExpansion
\right)  =L\left(
%TCIMACRO{\QATOP{\QATOP{\alpha_{1}}{0}}{\QATOP{0}{0}}}%
%BeginExpansion
\genfrac{}{}{0pt}{}{\genfrac{}{}{0pt}{}{\alpha_{1}}{0}%
}{\genfrac{}{}{0pt}{}{0}{0}}%
%EndExpansion
\right)
\end{equation}
We write out the $L=J$ in Eq.(\ref{GL}) more explicitly in terms of
$\frac{s_{d}}{4}\times\frac{s_{d}}{4}$ blocks by using the gamma matrices in
footnote (\ref{gamms}), and we obtain
\begin{align}
\mu &  =\left(
\begin{array}
[c]{c}%
L^{+^{\prime}-^{\prime}}+L^{r0}\rho_{r}+iL^{12}+\frac{1}{2}L_{rs}\rho^{rs}\\
-L^{01}-iL^{02}-\left(  L^{r1}+iL^{r2}\right)  \rho_{r}%
\end{array}
\right)  hu,\;\;\label{mu}\\
\lambda &  =\left(
\begin{array}
[c]{c}%
L^{+^{\prime}0}+L^{+^{\prime}r}~\rho_{r}\\
L^{+^{\prime}1}+iL^{+^{\prime}2}%
\end{array}
\right)  i\sqrt{2}hu.\; \label{lambd}%
\end{align}
Here $\alpha_{1}=hu$ is written as a product of a unitary matrix $u$ and a
Hermitian matrix $h.$ The matrix $u$ is an arbitrary $\frac{s_{d}}{4}%
\times\frac{s_{d}}{4}$ unitary matrix that belongs to the gauge group that
acts on the twistors. The generator of this gauge group is the constraint
$\left(  \bar{Z}Z\right)  _{a}^{~b}=0.$ The matrix $h$ is determined by
insuring $\lambda\bar{\lambda}=\frac{1}{\sqrt{2}}L^{+^{\prime}\mu}\gamma_{\mu
}$ that was established in Eq.(\ref{lambda}) and is given by
\begin{equation}
h=2^{-3/4}\left(  L^{+^{\prime}0}+L^{+^{\prime}r}~\rho_{r}\right)  ^{-1/2}.
\label{h}%
\end{equation}

To summarize, in Eqs.(\ref{mu}-\ref{h}) we have given a more fundamental form
of the twistor transform between $Z_{A}^{~a}$ and the $\left(  d+2\right)  $
dimensional phase space $\left(  X^{M},P^{M}\right)  .$ The formulas are
Sp$\left(  2,R\right)  $ gauge invariant since $\left(  X,P\right)  $ appear
only in the form $L^{MN}=X^{M}P^{N}-X^{N}P^{M}.$ These transforms are
consistent with Eqs.(\ref{tttransf},\ref{mulambda},\ref{null}) which somewhat
resemble the more traditional form of the twistor transform.

Note that $Z\left(  L\right)  $ written in terms of $L^{MN}$ depends only on
$2\left(  d-1\right)  $ independent combinations of the $L^{MN},$ since the
$L^{MN}$ obey the constraints $L^{MN}L_{NK}=0$ that follow from $X^{2}%
=P^{2}=X\cdot P=0$ in the physical sector. From these we derive the explicit
relations%
\[
\left(  L^{+^{\prime}-^{\prime}}\right)  ^{2}=-L^{+^{\prime}\mu}L_{~\mu
}^{-^{\prime}},\;L^{\mu\nu}=\frac{L^{+^{\prime}\mu}L^{-^{\prime}\nu
}-L^{+^{\prime}\nu}L^{-^{\prime}\mu}}{L^{+^{\prime}-^{\prime}}},\;L^{+^{\prime
}\mu}L_{~\mu}^{+^{\prime}}=0=L^{-^{\prime}\mu}L_{~\mu}^{-^{\prime}}.
\]
So, all $L^{MN}$ are then written only in terms of the $2d$ vector components
$L^{\pm^{\prime}\mu},$ but those are lightlike vectors and therefore contain
only $2\left(  d-1\right)  $ independent degrees of freedom.

Both the 2T particle and the corresponding twistors are SO$\left(  d,2\right)
$ covariant descriptions, the first is in terms of vectors $X_{i}^{M}=\left(
X^{M},P^{M}\right)  $ and the second is in terms of spinors $Z_{A}^{~a}$ of
SO$\left(  d,2\right)  $. This covariance is achieved by having gauge
symmetries in both versions, in the first case the gauge symmetry is applied
on the $i$ index of $X_{i}^{M}$ and in the second case the gauge symmetry is
applied on the $a$ index of $Z_{A}^{~a}.$

To account for the gauge symmetry and the constraint (\ref{ZZconstraint}) we
may derive them from a twistor action principle. This is done by rewriting the
twistor action in Eq.(\ref{Stw}) in a form similar to Eq.(\ref{action})
\begin{equation}
S_{twistor}=\frac{4}{s_{d}}\int d\tau~Tr\left[  \left(  i\bar{Z}DZ\right)
-h_{d}V\right]  ,\;\;\left(  DZ_{A}\right)  ^{a}\equiv\frac{\partial
Z_{A}^{~a}}{\partial\tau}-iZ_{A}^{~b}V_{b}^{~a}. \label{stwistor}%
\end{equation}
The equation of motion of the $\frac{s_{d}}{4}\times\frac{s_{d}}{4}$ matrix
gauge field $V_{b}^{~a}$ generates the constraint $\left(  \bar{Z}Z\right)
_{a}^{~b}-h_{d}\delta_{a}^{~b}=0$ where $h_{d}$ is chosen to make the matrix
traceless or with trace depending on the number of dimensions $d$ (see e.g.
the counting of degrees of freedom for $d=5,6$ following
Eq.(\ref{ZZconstraint})). Equivalently, the matrix $V$ itself can be taken as
traceless or with trace depending on the dimension, and this will result in
the same constraint. Once this constraint is satisfied this action reduces to
the previous one in Eq.(\ref{Stw}). For $d=3,4,5,6$ this is the full action
principle in terms of twistors since there are no other conditions (other than
reality or pseudo-reality in some dimensions) as discussed following
Eq.(\ref{ZZconstraint}). However, for $d\geq7$ some more conditions on a
general complex or (pseudo)real $Z$ are required to make it satisfy also the
basic relation Eq.(\ref{basicrelation}) or its equivalent null conditions in
Eq.(\ref{null}). Although we have given the full twistor transform for any
dimension, we have so far given the full action principle only for $d\leq6.$

Up to now we have described the twistor transform for the \textquotedblleft
free\textquotedblright\ 2T particle in $d+2$ dimensions. But from here it is
an easy step to obtain the twistor transform for an assortment of non-trivial
particle dynamics in 1T-physics. Having established the transform between
twistors and the Sp$\left(  2,R\right)  $ doublets $\left(  X^{M}%
,P^{M}\right)  $ in $\left(  d+2\right)  $ dimensions, we can now make
Sp$\left(  2,R\right)  $ gauge choices to produce various dynamical systems in
$\left(  d-1\right)  +1$ dimensions, in the physical sector that satisfies
$X^{2}=P^{2}=X\cdot P=0$. Examples that occur in this paper are the massless
particle in $d$ dimensions of Eq.(\ref{massless1},\ref{massless2}), or the
massive particle in $d$ dimensions of Eq.(\ref{massive2x},\ref{massive2p}).
Other Sp$\left(  2,R\right)  $ gauge choices that include interacting and
curved background cases are found in \cite{2treviews}\cite{2tHandAdS}. By
inserting the gauge choice for $\left(  X^{M},P^{M}\right)  $ into
Eqs.(\ref{tttransf},\ref{mulambda}) we obtain the corresponding twistors in
$d$ dimensions, such as the twistors for the massless particle of
Eq.(\ref{penrose}), or the massive particle in of Eq.(\ref{twistormassive}).
For more examples see \cite{twistorBP1} where the computations for the twistor
transform were done explicitly in $d=4,$ but the same explicit formulas also
apply in $d$ dimensions by inserting the corresponding gamma matrices in $d$
dimensions, as given in the expressions above.

\subsection{Geometry: twistors as the coset SO$\left(  d,2\right)
/$H$_{\Gamma}$ \label{coset}}

A geometric view of twistors in $d$ dimensions can also be given in the form
of a coset as follows. The starting point for twistors was the twistor gauge
of 2T-physics in Eq.(\ref{Stw}) which involved the group element $g\left(
\tau\right)  $ in the spinor representation of SO$\left(  d,2\right)  $ and
the special matrix $\Gamma$ in Eq.(\ref{g--}). The action and its SO$\left(
d,2\right)  $ global symmetry charge (on the left side of $g)$ have the form%
\begin{equation}
S\left(  g\right)  =\frac{4}{s_{d}}\int d\tau~Tr\left(  ig^{-1}\partial_{\tau
}g\Gamma\right)  ,\;\;\left(  J\right)  _{A}^{B}=\left(  g\Gamma
g^{-1}\right)  _{A}^{B}.\; \label{Sg}%
\end{equation}
The action is like a sigma model, but it is linear instead of being quadratic
in the Cartan connection $ig^{-1}\partial_{\tau}g,$ and has the special
insertion $\Gamma$ on the right side of $g.$ The insertion $\Gamma$ determines
important properties of this action. The equation of motion for $g $ is
$\left[  \Gamma,g^{-1}\partial_{\tau}g\right]  =0.$ Using this one can show
that the global current is conserved $\partial_{\tau}\left(  g\Gamma
g^{-1}\right)  =0,$ as expected from Noether's theorem.

We recall that the current $J_{A}^{~B}$ is gauge invariant and contains all
the physical information of the theory as seen in the previous sections. In
particular the current satisfies $J^{2}=\left(  g\Gamma g^{-1}\right)  \left(
g\Gamma g^{-1}\right)  =\left(  g\Gamma^{2}g^{-1}\right)  =0$ at the classical
level, which is consistent with the covariant approach in section
(\ref{ggroup}). This property of the current captures all the essential
aspects of the physical sector at the classical level.

Now, let us determine the independent degrees of freedom that contribute to
the current. We will find that there are precisely 2$\left(  d-1\right)  $
degrees of freedom, precisely equal to the number of physical degrees of
freedom. Since $g^{-1}\Gamma g$ is a SO$\left(  d,2\right)  $ transformation
applied on a generator $\Gamma$ in the algebra of SO$\left(  d,2\right)  ,$ we
can eliminate from $g\left(  \tau\right)  $ the subgroup H$_{\Gamma}$ that
leaves $\Gamma$ invariant, and keep only the coset degrees of freedom in
SO$\left(  d,2\right)  /$H$_{\Gamma}.$ To do this, we can decompose $g\left(
\tau\right)  =T_{\Gamma}\left(  \tau\right)  H_{\Gamma}\left(  \tau\right)  $
and write $g\Gamma g^{-1}=T_{\Gamma}\Gamma T_{\Gamma}^{-1}$ since by
definition $H_{\Gamma}\Gamma H_{\Gamma}^{-1}=\Gamma.$ Here $H_{\Gamma}%
=\exp\left(  h_{\Gamma}\right)  $ and $T_{\Gamma}=\exp\left(  t_{\Gamma
}\right)  ,$ where $h_{\Gamma}$ ($t_{\Gamma}$) is a linear combination of all
the SO$\left(  d,2\right)  $ generators $\Gamma^{MN}$ that commute (do not
commute) with $\Gamma,$ i.e. $\left[  h_{\Gamma},\Gamma\right]  =0$ and
$\left[  t_{\Gamma},\Gamma\right]  \neq0.$

To characterize the sets of generators $\left(  h_{\Gamma},t_{\Gamma}\right)
$ consider the decomposition of SO$\left(  d,2\right)  $ with respect to the
SO$\left(  d-2\right)  \times$SO$\left(  2,2\right)  $ subgroup. We can write
$J^{MN}=J^{ij}\oplus J^{\mu\nu}\oplus J^{\mu i}$ where $i=1,2,\cdots,\left(
d-2\right)  $ spans the SO$\left(  d-2\right)  $ basis and $\mu=+^{\prime
},-^{\prime},+,-$ (or $0^{\prime},0,1^{\prime},1$) spans the SO$\left(
2,2\right)  $ basis. Furthermore we decompose SO$\left(  2,2\right)
=$SL$\left(  2,R\right)  _{+}\times$SL$\left(  2,R\right)  _{-}$ and note that
each $\mu$ index is in the $\left(  \frac{1}{2},\frac{1}{2}\right)  $
representation of SL$\left(  2,R\right)  _{+}\times$SL$\left(  2,R\right)
_{-}. $ The SL$\left(  2,R\right)  _{+}\times$SL$\left(  2,R\right)  _{-}$
generators can be identified explicitly as
\begin{align}
\text{SL}\left(  2,R\right)  _{+}  &  :\;\frac{1}{2}\left(  J^{+^{\prime
}-^{\prime}}+J^{+-}\right)  ,~J^{+^{\prime}+},~J^{-^{\prime}-}\\
\text{SL}\left(  2,R\right)  _{-}  &  :\;\frac{1}{2}\left(  J^{+^{\prime
}-^{\prime}}-J^{+-}\right)  ,~J^{+^{\prime}-},~J^{-^{\prime}+}%
\end{align}
From the general SO$\left(  d,2\right)  $ commutation rules
\begin{equation}
\left[  J^{MN},J^{KL}\right]  =i\left[  \left(  J^{ML}\eta^{NK}-\left(
M\leftrightarrow N\right)  \right)  -\left(  K\leftrightarrow L\right)
\right]  ,
\end{equation}
with $\eta^{+^{\prime}-^{\prime}}=\eta^{+-}=-1$ and $\eta^{ij}=\delta^{ij},$
it is easy to verify that these indeed form the SL$\left(  2,R\right)
_{+}\times$SL$\left(  2,R\right)  _{-}$ algebra. Under commutation with the
generator $\frac{1}{2}\left(  J^{+^{\prime}-^{\prime}}+J^{+-}\right)  $ each
SO$\left(  d,2\right)  $ generator has a definite charge $0,\pm\frac{1}{2}%
,\pm1.$ We list the generators according to those charges as follows%
\begin{equation}%
\begin{tabular}
[c]{ccccc}%
$-1$ & $-\frac{1}{2}$ & $0$ & $+\frac{1}{2}$ & $+1$\\
$J^{-^{\prime}-}$ & $\left(
\begin{array}
[c]{c}%
J^{-i}\\
J^{-^{\prime}i}%
\end{array}
\right)  $ & $\left.
\begin{array}
[c]{c}%
J^{ij},\;\frac{J^{+^{\prime}-^{\prime}}+J^{+-}}{2}\\
\left(  J^{-^{\prime}+},\frac{J^{+^{\prime}-^{\prime}}-J^{+-}}{2}%
,J^{+^{\prime}-}\right)
\end{array}
\right.  $ & $\left(
\begin{array}
[c]{c}%
J^{+^{\prime}i}\\
J^{+i}%
\end{array}
\right)  $ & $J^{+^{\prime}+}$%
\end{tabular}
\end{equation}
These charges are conserved additively in the general commutation rules
$\left[  J^{MN},J^{KL}\right]  =\cdots$ given above. Furthermore the charge
$\pm\frac{1}{2}$ generators form doublets under SL$\left(  2,R\right)  _{-}$
and vectors under SO$\left(  d-2\right)  $ as indicated, while the charge
$\pm1$ generators are singlets under both. From this structure of the
commutation rules $\left[  J^{MN},J^{KL}\right]  =\cdots$ we easily see that
the generators that commute or do not commute with $\Gamma\sim\Gamma
^{-^{\prime}-}$ (or $J^{-^{\prime}-}$) are%
\begin{align}
h_{\Gamma}  &  :\;J^{-^{\prime}-},~\left(
\begin{array}
[c]{c}%
J^{-i}\\
J^{-^{\prime}i}%
\end{array}
\right)  ,\left.
\begin{array}
[c]{c}%
J^{ij}\;\\
\left(  J^{-^{\prime}+},\frac{J^{+^{\prime}-^{\prime}}-J^{+-}}{2}%
,J^{+^{\prime}-}\right)
\end{array}
\right. \\
t_{\Gamma}  &  :\;\frac{1}{2}\left(  J^{+^{\prime}-^{\prime}}+J^{+-}\right)
,\left(
\begin{array}
[c]{c}%
J^{+^{\prime}i}\\
J^{+i}%
\end{array}
\right)  ,J^{+^{\prime}+}%
\end{align}
We note that each set forms a subalgebra $\left[  h_{\Gamma},h_{\Gamma
}\right]  \sim h_{\Gamma},\;\left[  t_{\Gamma},t_{\Gamma}\right]  \sim
t_{\Gamma}$ while $\left[  h_{\Gamma},t_{\Gamma}\right]  \sim h_{\Gamma
}+t_{\Gamma}.$ In particular, within $t_{\Gamma}$ the generator $\frac{1}%
{2}\left(  J^{+^{\prime}-^{\prime}}+J^{+-}\right)  $ forms a U$\left(
1\right)  $ subgroup and classifies the others according to their charges
above, while the remaining coset $T_{\Gamma}/U\left(  1\right)  $ forms an
algebra similar to the Heisenberg algebra, $\left[  J^{+^{\prime}i}%
,J^{+j}\right]  =i\delta^{ij}J^{+^{\prime}+},$ since $J^{+^{\prime}+}$
commutes with both $J^{+^{\prime}i},J^{+j}.$ From this we conclude that the
general $T_{\Gamma}=\exp\left(  t_{\Gamma}\right)  $ can be parametrized as
follows%
\begin{equation}
T_{\Gamma}=\exp\left(  t_{\Gamma}\right)  =\exp\left(  \frac{1}{2}\left(
\Gamma^{+^{\prime}-^{\prime}}+\Gamma^{+-}\right)  \omega\left(  \tau\right)
\right)  \exp\left(  \Gamma^{+^{\prime}+}z\left(  \tau\right)  \right)
\exp\left(  \Gamma^{+i}k_{i}\left(  \tau\right)  +\Gamma^{+^{\prime}i}%
q_{i}\left(  \tau\right)  \right)
\end{equation}

The number of parameters in this coset is precisely $2\left(  d-1\right)  ,$
which is the same as the number of physical degrees of freedom. If we insert
the explicit set of gamma matrices $\Gamma^{MN}$ given in footnote
(\ref{gamms}) into this expression, we can write $T_{\Gamma}$ in the form of
$\frac{s_{d}}{4}\times\frac{s_{d}}{4}$ similar to $\Gamma$. Then, as seen from
Eq.(\ref{Stw}), the first block of columns that forms an $s_{d}\times
\frac{s_{d}}{4}$ rectangular matrix is the twistor $Z_{A}^{~a}\left(
t_{\Gamma}\right)  $ now written in terms of only the $2\left(  d-1\right)  $
parameters of the coset $t_{\Gamma}\in$SO$\left(  d,2\right)  /$H$_{\Gamma}$.

This result is of course in agreement with the form of the Sp$\left(
2,R\right)  $ invariant $Z_{A}^{~a}\left(  L\right)  $ of Eqs.(\ref{mu}%
,\ref{lambd}), which can also be written only in terms of $2\left(
d-1\right)  $ parameters as detailed in Sec.(\ref{2dm1}). We now understand
that there is a close relationship with the geometric interpretation as a coset.

\section{D-branes and twistors}

In the discussion following the twistor action in Eq.(\ref{stwistor}) we
explained that more conditions are needed on $Z$ in order to obtain the proper
twistor that is equivalent to the phase space of a particle when $d\geq7.$
What if those conditions are never imposed? What would then be the content of
$Z_{A}^{~a}$ ? We find that the extra degrees of freedom can be interpreted as
collective coordinates of D-branes.

To see this let us consider the properties of the rectangular matrix
$Z_{A}^{~a}$ that follow from the action in Eq.(\ref{stwistor}). The gauge
group acts on the right side and there is a global symmetry with conserved
charges $J$ that acts on the left side$.$ These properties are summarized by
the equations%
\begin{equation}
\bar{Z}Z=0,\;J=Z\bar{Z}-trace.
\end{equation}
Thus, the global current $J$ is a $s_{d}\times s_{d}$ matrix in the
fundamental representation of the global group $G$ that can be expanded in a
complete set of SO$\left(  d,2\right)  $ gamma matrices as follows
\begin{equation}
J=Z\bar{Z}=J_{0}+\Gamma^{M}J_{M}+\frac{1}{2}\Gamma^{MN}J_{MN}+\frac{1}%
{3!}\Gamma^{MNK}J_{MNK}+\cdots\label{ggg}%
\end{equation}
There are as many terms $\Gamma^{M_{1}\cdots M_{n}}$ as necessary to span all
the generators of some group $G$ whose fundamental representation has the same
dimension $s_{d}$ of the SO$\left(  d,2\right)  $ spinor. If the group is
constrained to be SO$\left(  d,2\right)  $ there is only the term proportional
to $\Gamma^{MN}$ in the expansion (\ref{ggg}), and then the group element $g$
is constructed by exponentiating the generators as in Eq.(\ref{g}). But if the
group is more general, then the exponent in Eq.(\ref{g}) contains all the
terms that appear in Eq.(\ref{ggg}).

Thus, with a more general $g$ represented as a $s_{d}\times s_{d}$ matrix, a
more general twistor $Z$ would emerge, with more degrees of freedom than the
particle phase space. This is achieved with an action that is written in the
form of Eq.(\ref{Sg}), which in turn is obtained by gauge fixing from the
2T-physics action in (\ref{S2T}), but by taking $g$ to be a group element not
just in SO$\left(  d,2\right)  ,$ but an element in the smallest group $G$
that contains SO$\left(  d,2\right)  $ in the spinor representation. The
parent 2T-physics theory in Eq.(\ref{S2T}) has an interaction term of the form
of $\frac{4}{s_{d}}Tr\left(  ig^{-1}\partial_{\tau}gL\right)  $ which is
unchanged. But we emphasize that now $L$ is proportional to only $\Gamma
^{MN},$ while $ig^{-1}\partial_{\tau}g$ has all the terms in Eq.(\ref{ggg}),
so $L$ couples to only the SO$\left(  d,2\right)  $ subgroup of $G.$ Then the
action in Eq.(\ref{S2T}) still has local Sp$\left(  2,R\right)  $ and
SO$\left(  d,2\right)  $ symmetries, but now it has global $G$ symmetry
instead of only global SO$\left(  d,2\right)  $ on the left side of $g.$

This generalization of the group element $g$ allows $Z$ to contain the extra
degrees of freedom. We emphasize that the spinor representation of SO$\left(
d,2\right)  ,$ whose dimension is $s_{d},$ must correspond to the fundamental
representation of $G.$ This requirement determines $G$ as we see in Table 1 below.

We already know that for $d=3,4,6$ the groups $G=$Sp$\left(  4,R\right)  $,
SU$\left(  2,2\right)  $, Spin$\left(  6,2\right)  $ respectively are exactly
equal to SO$\left(  d,2\right)  $ in the spinor representation. Therefore for
these cases there are no other terms in Eq.(\ref{ggg}) other than $\Gamma
^{MN},$ provided some (pseudo)reality conditions are imposed as given
following Table 1 below. Extra terms usually appear for $d\geq7.$

As an example of what terms appear, consider SO$\left(  7,2\right)  $ for
$d=7$. The spinor representation has dimension 16. The smallest group with a
$16$ dimensional fundamental representation is SO$^{\ast}\left(  16\right)  $,
where the $^{\ast}$ indicates the appropriate analytic continuation that
contains SO$\left(  7,2\right)  $ as a subgroup. The number of generators of
SO$^{\ast}\left(  16\right)  $ is $\frac{16\cdot15}{2}=\allowbreak120.$ The
number of generators represented by the 16$\times16$ gamma matrices is
$\Gamma^{M}\rightarrow9$, $\Gamma^{MN}\rightarrow\frac{9\cdot8}{2}=36$,
$\Gamma^{MNK}\rightarrow\frac{9\cdot8\cdot7}{1\cdot2\cdot3}=\allowbreak84$,
$\Gamma^{MNKL}\rightarrow\frac{9\cdot8\cdot7\cdot6}{1\cdot2\cdot3\cdot
4}=\allowbreak126.$ We see that the 120 generators of SO$^{\ast}\left(
16\right)  $ are represented by $\frac{1}{2}\Gamma^{MN}L_{MN}+\frac{1}%
{3!}\Gamma^{MNK}L_{MNK}.$ Therefore the more general twistor in $d=7$ has the
expansion%
\begin{equation}
d=7:\;J=Z\bar{Z}=\frac{1}{2}L^{MN}\Gamma_{MN}+\frac{1}{3!}L^{MNK}\Gamma_{MNK}.
\end{equation}
If we impose additional conditions on $Z$ as in the previous sections, then we
eliminate the term $L_{MNK}=0,$ and the remaining $L_{MN}$ necessarily
satisfies all the conditions of section (\ref{sttransform}) since they all
followed from $\bar{Z}Z=0$ that was imposed by the gauge symmetry H.

However, if we do not impose the conditions of section (\ref{sttransform})
then we can interpret the degrees of freedom $L_{MNK}$ as D-brane degrees of
freedom. To see this, consider the smallest extended super-\textit{conformal}
algebra that contains spin$\left(  7,2\right)  \subset$SO$^{\ast}\left(
16\right)  $. This is OSp$\left(  16|2\right)  .$ Its two supercharges satisfy
$\left\{  Q_{A}^{i},Q_{B}^{j}\right\}  =\varepsilon^{ij}\left[  \frac{1}%
{2}\left(  \Gamma^{MN}\right)  _{AB}~L_{MN}+\frac{1}{3!}\left(  \Gamma
^{MNK}\right)  _{AB}~L_{MNK}\right]  +q^{ij}C_{AB},$ with $i=1,2$ labelling
the Sp$\left(  2\right)  ,$ and $C_{AB},q^{ij}$ both symmetric. The usual
$d=7$ Poincar\'{e} super-algebra is a subalgebra obtained from the above by
decomposing the SO(7,2)$\rightarrow$SO(6,1)$\times$SO(1,1) for spinors (16
=8$_{+}$+8$_{-}$ ) as $A=\alpha_{+}\oplus\alpha_{-}$ and vectors as
$M=\pm^{\prime},\mu,$ and keeping only the operators $Q_{\alpha+}^{i}$ and all
the $L^{MN},L^{MNK}$ with a single $+^{\prime}$ as follows%
\begin{equation}
\left\{  Q_{\alpha+}^{1},Q_{\beta+}^{2}\right\}  =L^{+^{\prime}\mu}\left(
\Gamma_{+^{\prime}\mu}\right)  _{\alpha\beta}+\frac{1}{2}L^{+^{\prime}\mu\nu
}\left(  \Gamma_{+^{\prime}\mu\nu}\right)  _{\alpha\beta}. \label{Dbrane}%
\end{equation}
In the massless particle gauge, $L^{+^{\prime}\mu}$ is the momentum $p^{\mu}$
and then $L^{+^{\prime}\mu\nu}$ are the \textit{commuting} D2-brane charges in
$d=7$ dimensions (like generalized momenta). The other components of $L_{MN},$
and $L_{MNK}$ are functions of phase space, including the particle as well as
D-brane canonical degrees of freedom, and do not generally commute among
themselves\footnote{The commutation rules of the $L^{M_{1}\cdots M_{n}}$ are
isomorphic to the commutation rules of the $\Gamma^{M_{1}\cdots M_{n}}.$ From
this we see that the D-brane charges $L^{+^{\prime}\mu\nu}$ commute among
themselves as well as with the momenta $p^{\mu}=L^{+^{\prime}\mu
}.\label{Dcharges}$}.

As another example consider SO$\left(  8,2\right)  $ for $d=8$. The two spinor
representations are 16,$\overline{16}$. The smallest groups with $16$
dimensional fundamental representations are SO$^{\ast}\left(  16\right)  $,
Sp$^{\ast}\left(  16\right)  $, SU$^{\ast}\left(  16\right)  .$ To decide
which is the smallest one that contains SO$\left(  8,2\right)  $ as a subgroup
we analyze the number of generators represented by the gamma matrices. The
number of generators of SO$^{\ast}\left(  16\right)  $ is $\frac{16\cdot15}%
{2}=\allowbreak120,$ for Sp$^{\ast}\left(  16\right)  $ is $\frac{16\cdot
17}{2}=\allowbreak136,$ and for SU$^{\ast}\left(  16\right)  $ is $\left(
16\right)  ^{2}-1=255.$ The number of generators represented by the gamma
matrices is $\Gamma^{M}\rightarrow10$, $\Gamma^{MN}\rightarrow\frac{10\cdot
9}{2}=\allowbreak45$, $\Gamma^{MNK}\rightarrow\frac{10\cdot9\cdot8}%
{1\cdot2\cdot3}=\allowbreak120$, $\Gamma^{MNKL}\rightarrow\frac{10\cdot
9\cdot8\cdot7}{1\cdot2\cdot3\cdot4}=210$, $\Gamma_{+}^{MNKLR}\rightarrow
\frac{1}{2}\frac{10\cdot9\cdot8\cdot7\cdot6}{1\cdot2\cdot3\cdot4\cdot
5}=\allowbreak126,$ where the last one is self dual (hence the extra factor of
$\frac{1}{2}$). The 45 generators of SO$\left(  8,2\right)  $ represented by
$\Gamma^{MN}$ must be included as one of the criteria in choosing the smallest
$G$. Then we see that there is no combination of gamma matrices that can be
used to construct SO$^{\ast}\left(  16\right)  $ and Sp$^{\ast}\left(
16\right)  $ and therefore we must take SU$^{\ast}\left(  16\right)  $ as the
smallest group that contains spin$\left(  8,2\right)  .$ The smallest
superconformal algebra is SU$\left(  16|1\right)  .$ The 255 generators of
SU$^{\ast}\left(  16\right)  $ that appear in $\left\{  Q_{A},\bar{Q}%
^{B}\right\}  $ are then represented by $\frac{1}{2}\Gamma^{MN}L_{MN}+\frac
{1}{4!}\Gamma^{MNKL}L_{MNKL}.$ The extra $L_{MNKL}$ lead to the D-brane
degrees of freedom. To see the content of D-brane \textit{commuting}%
$^{\ref{Dcharges}}$ charges we must decompose SO$\left(  8,2\right)  $ to
SO$\left(  7,1\right)  $ by $M=\pm^{\prime},\mu$ and identify the D-brane
commuting charges as the D3-brane $Z^{+^{\prime}\mu\nu\lambda}$ in $8$ dimensions.%

\begin{gather*}%
\begin{tabular}
[c]{|l|l|l|l|l|l|l|}\hline
d & Spin$\left(  d,2\right)  $ & {\small spinor} & G & G$_{\text{super}%
}\left(  N\right)  $ & {\small generators of G in Spin}$\left(  d,2\right)
${\small basis} & $%
%TCIMACRO{\QATOP{\text{contained}}{\text{in product}}}%
%BeginExpansion
\genfrac{}{}{0pt}{}{\text{contained}}{\text{in product}}%
%EndExpansion
$\\\hline
3 & Spin$\left(  3,2\right)  $ & 4 & Sp$\left(  4,R\right)  $ & OSp$\left(
N|4\right)  $ & $\Gamma^{MN}$ $_{10}$ & $\left(  {\small 4\times4}\right)
_{s}$\\\hline
4 & Spin$\left(  4,2\right)  $ & 4$,\bar{4}$ & SU$\left(  2,2\right)  $ &
SU$\left(  2,2|N\right)  $ & $\Gamma^{MN}$ $_{15}$ & ${\small 4\times\bar{4}}%
$\\\hline
5 & Spin$\left(  5,2\right)  $ & 8$_{+}$ & $%
%TCIMACRO{\QATOP{\text{spin}^{\ast}\text{(7)~}}{\text{SO}^{\ast}\left(
%8\right)  ~~}}%
%BeginExpansion
\genfrac{}{}{0pt}{}{\text{spin}^{\ast}\text{(7)~}}{\text{SO}^{\ast}\left(
8\right)  ~~}%
%EndExpansion
$ & $%
%TCIMACRO{\QATOP{\text{F(4)~~~~~~~~~~}}{\text{OSp}\left(  8|2N\right)  ~}}%
%BeginExpansion
\genfrac{}{}{0pt}{}{\text{F(4)~~~~~~~~~~}}{\text{OSp}\left(  8|2N\right)  ~}%
%EndExpansion
$ & $%
%TCIMACRO{\QATOP{\Gamma^{MN}\text{~}_{21}\text{~~~~~~~~~~~~~~}}{\Gamma
%^{MN}~_{21}~\oplus~~\Gamma^{M}~_{7}}}%
%BeginExpansion
\genfrac{}{}{0pt}{}{\Gamma^{MN}\text{~}_{21}\text{~~~~~~~~~~~~~~}}{\Gamma
^{MN}~_{21}~\oplus~~\Gamma^{M}~_{7}}%
%EndExpansion
$ & $\left(  {\small 8\times8}\right)  _{a}$\\\hline
6 & Spin$\left(  6,2\right)  $ & 8$_{+}$ & SO$^{\ast}\left(  8\right)  $ &
OSp$\left(  8|2N\right)  $ & $\Gamma^{MN}$ $_{28}$ & $\left(  {\small 8\times
8}\right)  _{a}$\\\hline
7 & Spin$\left(  7,2\right)  $ & 16 & SO$^{\ast}\left(  16\right)  $ &
OSp$\left(  16|2N\right)  $ & $\Gamma^{MN}$ $_{36}$ $\oplus$ $\Gamma^{MNK}$
$_{84}$ & $\left(  1{\small 6\times16}\right)  _{a}$\\\hline
8 & Spin$\left(  8,2\right)  $ & {\small 16}$,\overline{{\small 16}}$ &
SU$^{\ast}\left(  16\right)  $ & SU$\left(  16|N\right)  $ & $\Gamma^{MN}$
$_{45}$ $\oplus$ $\Gamma^{MNKL}$ $_{210}$ & ${\small 16\times~}\overline
{{\small 16}}$\\\hline
9 & Spin$\left(  9,2\right)  $ & 32 & Sp$^{\ast}\left(  32\right)  $ &
OSp$\left(  N|32\right)  $ & $\Gamma^{MN}$ $_{55}$ $\oplus$ $\Gamma^{M}$
$_{11}$ $\oplus$ $\Gamma^{M_{1}\cdots M_{5}}$ $_{462}$ & $\left(
{\small 32\times32}\right)  _{s}$\\\hline
10 & Spin$\left(  10,2\right)  $ & 32$_{+}$ & Sp$^{\ast}\left(  32\right)  $ &
OSp$\left(  N|32\right)  $ & $\Gamma^{MN}$ $_{66}$ $\oplus$ $\Gamma_{+}%
^{M_{1}\cdots M_{6}}$ $_{462}$ & $\left(  {\small 32\times32}\right)  _{s}%
$\\\hline
11 & Spin$\left(  11,2\right)  $ & 64 & Sp$^{\ast}\left(  64\right)  $ &
OSp$\left(  N|64\right)  $ & $\Gamma^{MN}$ $_{78}$ $\oplus$ $\Gamma^{MNK}$
$_{286}$ $\oplus$ $\Gamma^{M_{1}\cdots M_{6}}$ $_{1716}$ & $\left(
{\small 64\times64}\right)  _{s}$\\\hline
12 & Spin$\left(  12,2\right)  $ & {\small 64,}$\overline{{\small 64}}$ &
SU$^{\ast}\left(  64\right)  $ & SU$\left(  64|N\right)  $ & $\Gamma^{MN}$
$_{91}$ $\oplus$ $\Gamma^{MNKL}$ $_{1001}$ $\oplus$ $\Gamma^{M_{1}\cdots
M_{6}}$ $_{3003}$ & ${\small 64\times~}\overline{{\small 64}}$\\\hline
\end{tabular}
\\
\text{{\small Table 1:} {\small Smallest} {\small groups} }{\small G}\text{
{\small and supergroups} }{\small G}_{\text{super}}\text{ {\small that contain
Spin}}\left(  {\small d,2}\right)  {\small ,}\text{ {\small and D-branes}.}%
\end{gather*}

In Table-1 we give a list of the smallest groups $G$ that contain spin$\left(
d,2\right)  $ for $3\leq d\leq12.$ We also include the smallest supergroup
$G_{\text{super}}$ that contains $G.$ We list the gamma matrix representation
of the generators of $G$, and their numbers as subscripts, as represented by
antisymmetrized products of gamma matrices $\Gamma^{M_{1}\cdots M_{n}}%
\equiv\frac{1}{n!}\left(  \Gamma^{M_{1}}\bar{\Gamma}^{M_{2}}\Gamma^{M_{3}%
}\cdots\Gamma^{M_{n}}\mp\text{permutations}\right)  $ in dimension $d+2$
labelled by $M.$ The last column gives information on whether the gamma
matrices occur in the symmetric or antisymmetric products of the spinors of
SO$\left(  d,2\right)  $, when both spinor indices $A,B$ are lowered or raised
in the form $\left(  \Gamma^{M_{1}\cdots M_{n}}\right)  _{AB}$ by using the
metric $C$ in spinor space. The phase space of the D-branes correspond to the
extra generators beyond $\Gamma^{MN}$ as explained in the examples above. In
the case of $d=5,$ one option is to keep the D0-brane associated with
$\Gamma^{M}\rightarrow\Gamma^{+^{\prime}},$ another option is to remove it
with the extra U$\left(  1\right)  $ gauge symmetry as discussed in the
counting done in section (\ref{sttransform}). The D-brane does not occur for
the supergroup F$\left(  4\right)  $ that can be used for the $d=5$
superparticle as described below.

Groups that are larger than the listed $G$ may be considered in our scheme in
every dimension (e.g. SU$\left(  8\right)  $ instead of SO$\left(  8\right)  $
in $d=6$, etc.). In that case the number of generators $\Gamma^{M_{1}\cdots
M_{n}}$ increases compared to the ones listed in the table for each $d.$
Furthermore the corresponding D-brane degrees of freedom also get included in
the model.

When $Z_{A}^{~a}$ is obtained from the group element $g$ through the relation
$Z\bar{Z}=g\Gamma g^{-1},$ with the group $G$ listed in the table above, then
$Z$ is real or pseudo-real when the group is SO or Sp and it is complex when
the group is SU. Given those properties, in general the quadratic $\left(
Z\bar{Z}\right)  _{A}^{~~B}$ contains just the gamma matrices listed above
which correspond to the generators of the group $G.$ If the (pseudo)reality
properties associated with $G$ are not obeyed by $Z$ then more D-brane terms
will appear generally in the expansion of $Z\bar{Z}$ as in Eq.(\ref{ggg}) as
compared to those on the table.

The case of $d=11$ is particularly interesting since it relates to M-theory as
follows. The corresponding twistors are spinors of Spin$(11,2)$ that are 64
dimensional. The smallest group is Sp$^{\ast}\left(  64\right)  $ whose
generators are represented by $\Gamma^{MN}$ $\left(  78\right)  $ +
$\Gamma^{MNK}$ $\left(  286\right)  $ + $\Gamma^{M_{1}\cdots M_{6}}$ $\left(
1716\right)  .$ To identify the commuting charges of D-branes we decompose
$M=\pm^{\prime},\mu$ and keep all the generators with a single $+^{\prime},$
as follows $L^{+^{\prime}\mu}$ $\oplus$ $L^{+^{\prime}\mu\nu}$ $\oplus$
$L^{+^{\prime}\mu_{1}\cdots\mu_{5}}.$ Here $L^{+^{\prime}\mu}$ is the momentum
in 11 dimensions and $L^{+^{\prime}\mu\nu},L^{+^{\prime}\mu_{1}\cdots\mu_{5}}$
are the D2-brane and D5-brane \textit{commuting}$^{\ref{Dcharges}}$ charges respectively.

Let us mention that the discussion above with the group $G$ can be directly
generalized to the supergroup $G_{\text{super}}\left(  N\right)  $ listed in
Table 1 (with some limits on $N$ as discussed below) by following
\cite{2ttwistor}. The 2T-physics action is still of the same form as $S\left(
X,P,g\right)  $ of Eq.(\ref{S2T}), but now we have a supergroup element $g$
and a supertrace coupling $\frac{4}{s_{d}}Tr\left(  ig^{-1}\partial_{\tau
}gL\right)  ,$ and the matrix $L$ is of the form $L=\frac{1}{4i}\left(
%TCIMACRO{\QATOP{\Gamma^{MN}}{0}}%
%BeginExpansion
\genfrac{}{}{0pt}{}{\Gamma^{MN}}{0}%
%EndExpansion%
%TCIMACRO{\QATOP{0}{0}}%
%BeginExpansion
\genfrac{}{}{0pt}{}{0}{0}%
%EndExpansion
\right)  L_{MN}$ where the gamma matrices couple to the bosonic subgroup $G$
as above. This 2T superparticle action reduces to the standard massless
superparticle action in the particle gauge for dimensions $d=3,4,5,6$
\cite{2ttwistor} with $N$ supersymmetries. It can also be gauge fixed to the
twistor gauge to give supertwistors that are equivalent to the super phase
space in those dimensions \cite{2tsuperstring}\cite{2tstringtwistors}. One can
go beyond those gauge choices and obtain a twistor description of many other
dual superparticle theories that give the super generalizations of the ones
studied recently in \cite{twistorBP1}.

The supergroup can be enlarged to have more fermionic generators, but keeping
$G$ as a bosonic subgroup. For example, for $d=4$ we may take SU$\left(
2,2|N\right)  $ instead of the smallest $N=1$ shown in the table. For physical
purposes the total number of real fermionic generators cannot exceed 64 (32
ordinary supercharges and 32 conformal supercharges). For example, for $d=4$
we can go as far as $N=8$, or $G_{\text{super}}=$SU$\left(  2,2|8\right)  $
which has 64 real fermionic parameters. Similarly, for $d=10$ we may take
OSp$\left(  1|32\right)  $ or OSp$\left(  2|32\right)  .$ In the more general
cases the coupling $L$ can be of the form of the previous paragraph
$L=\frac{1}{4i}\left(
%TCIMACRO{\QATOP{\Gamma^{MN}}{0}}%
%BeginExpansion
\genfrac{}{}{0pt}{}{\Gamma^{MN}}{0}%
%EndExpansion%
%TCIMACRO{\QATOP{0}{0}}%
%BeginExpansion
\genfrac{}{}{0pt}{}{0}{0}%
%EndExpansion
\right)  L_{MN}.$

The model can also be generalized by adding $d^{\prime}$ more dimensions
$\left(  X^{I},P^{I}\right)  $ in addition to the $d+2$ dimensions $\left(
X^{M},P^{M}\right)  $, but keeping the same $g\in G_{\text{super}}.$ The
generalized action has the form \cite{2tAdSs}\cite{2tsuperstring}%
\cite{2tstringtwistors}\cite{chinaLect}%
\begin{equation}
S_{2T}\left(  \hat{X},\hat{P},g\right)  =\int d\tau\left[  \frac{1}%
{2}\varepsilon^{ij}\partial_{\tau}\hat{X}_{i}\cdot\hat{X}_{j}-\frac{1}%
{2}A^{ij}\hat{X}_{i}\cdot\hat{X}_{j}+\frac{4}{s_{d}}Str\left(  ig^{-1}%
\partial_{\tau}g\hat{L}\right)  \right]
\end{equation}
where $\hat{X}^{\hat{M}}=\left(  X^{M},X^{I}\right)  ,$ $\hat{P}^{\hat{M}%
}=\left(  P^{M},P^{I}\right)  ,$ and we now take the more general coupling
$\hat{L}=\frac{1}{4i}\left(
%TCIMACRO{\QATOP{\Gamma^{MN}}{0}}%
%BeginExpansion
\genfrac{}{}{0pt}{}{\Gamma^{MN}}{0}%
%EndExpansion%
%TCIMACRO{\QATOP{0}{0}}%
%BeginExpansion
\genfrac{}{}{0pt}{}{0}{0}%
%EndExpansion
\right)  L_{MN}+\frac{\alpha}{4i}\left(
%TCIMACRO{\QATOP{0}{0}}%
%BeginExpansion
\genfrac{}{}{0pt}{}{0}{0}%
%EndExpansion%
%TCIMACRO{\QATOP{0}{\Gamma^{IJ}}}%
%BeginExpansion
\genfrac{}{}{0pt}{}{0}{\Gamma^{IJ}}%
%EndExpansion
\right)  L_{IJ}$. The fixed parameter $\alpha=\frac{s_{d}}{s_{d^{\prime}}}$ is
determined by local bosonic and fermionic symmetries in this action, to be the
ratio of the spinor dimensions of SO$\left(  d+2\right)  $ and SO$\left(
d^{\prime}\right)  .$ In this latter scheme we obtain interesting cases, such
as supertwistors with some compactified subspaces, without D-branes. For
example supertwistors for AdS$_{4}\times$S$^{7},$ AdS$_{5}\times$S$^{5},$
AdS$_{7}\times$S$^{4}$ with a of total 10 or 11 dimensions emerge
\cite{2tsuperstring}\cite{2tstringtwistors}\cite{chinaLect} by using
supergroups with only 32 real fermions, namely $G_{\text{super}}=$OSp$\left(
8|4^{\ast}\right)  $, SU$\left(  2,2|4\right)  $, OSp$\left(  8^{\ast
}|4\right)  $ respectively.

This analysis taken to the maximum allowed number of supersymmetries and the
maximum number of dimensions leads to $d=11$ with OSp$\left(  1|64\right)  $
as the hidden global supersymmetry for M-theory, and suggests that the
extended supertwistors may well play a role in a nice description of M-theory.
This can be studied through the toy M-model \cite{2ttoyM} that has the action
$S\left(  X,P,g\right)  $ of Eq.(\ref{S2T}) with the group OSp$\left(
1|64\right)  ,$ and includes the D2 and D5-branes. The supergroup OSp$\left(
1|64\right)  $ is motivated by other considerations as well \cite{Stheory}%
\cite{liftM}\cite{west}. In other approaches to twistors in D=11
\cite{azcarraga2}\cite{azcarraga3}, twistors in the fundamental representation
of OSp$(1|64)$ were used in \cite{azcarraga3} for the formulation of a
superstring action in an extended D=11 superspace.

The twistor action (\ref{stwistor}) is a gauge fixed form of the 2T-physics
action $S\left(  X,P,g\right)  $ of Eq.(\ref{S2T}) for a general $g$. We can
play the game of gauge fixing the local symmetries Sp$\left(  2,R\right)  $
and SO$\left(  d,2\right)  $ of $S\left(  X,P,g\right)  $ in many possible
ways and derive a multitude of 1T-physics systems with a rich web of dualities
among them. D-branes are included in this web of dualities. Then twistors can
be shown to unify many dual theories including D-branes. It would be
interesting to pursue this line of reasoning in more detail.

\section{Discussion}

In this paper we gave the general twistor transform that maps twistor space to
phase space in $d$ dimensions with one time. The general transform can be
specialized to a variety of special dynamical particle systems that include
particles with or without mass, relativistic or nonrelativistic, in flat or
curved spaces, interacting or non-interacting. Thus, the scope of our formulas
is much larger than the traditional twistor transform. The special cases of
phase space described by the same twistor are those that can be derived by
gauge fixing the parent unifying theory in 2T-physics. Thus, either the
twistor description or the vector SO$\left(  d,2\right)  $ description in
2T-physics provide a unification of those 1T-physics systems and establishes a
duality relationship between them.

To our knowledge this is the first time that twistors have been successfully
defined generally in $d$ dimensions. We have insured that our twistor
transform is fully equivalent to particle phase space for all dimensions. If
we specialize to the phase space of massless particles only, then our result
agrees with twistors that were previously defined for $d\leq6$ in another
approach \cite{cederwal}. Even for $d\leq6$ our twistors for the phase spaces
other than the massless particle are all new structures. For a rather
different approach to twistors for massive particles, which uses double the
number of twistors compared to our formulas and only in $d=4$, see
\cite{penrose2},\cite{perjes}-\cite{azcarraga}.

Beyond twistors for particles, we have also defined twistors for a phase space
that includes also D-brane degrees of freedom. Including the D-branes may lead
to some interesting applications of twistors, in particular for M-theory. The
twistor action principle in Eq.(\ref{Stw}) applies generally to twistors
including D-branes for a generally complex $Z$ in every dimension. If
(pseudo)reality conditions are imposed on $Z$ as mentioned after the table in
the previous section, then for $d\leq6$ we obtain only the particle phase
space out of the twistor. In $d\geq7$ there are automatically D-branes even
with the (pseudo) reality conditions. However, if extra constraints are
applied on $Z,$ as detailed in section (\ref{sttransform}), then again the
degrees of freedom in $Z$ are thinned down to only the particle phase space
without D-branes in every dimension.

Quantization of twistors for any spin in four dimensions was discussed in
section (\ref{twistd4}). Here we suggested a free field theory in twistor
space that describes any spinning particle. This could also lead to some
interesting applications of twistors in field theory with interactions in four dimensions.

Generalizations of our results in many directions are possible. Some of these
are already briefly described in recent papers \cite{2tsuperstring}%
\cite{2tstringtwistors}, such as twistors for spinning particles,
supertwistors in various dimensions, including compactified dimensions, and
supertwistors for supersymmetric AdS$_{5}\times$S$^{5},$ AdS$_{4}\times$%
S$^{7},$ AdS$_{7}\times$S$^{4}.$ We plan to give details of those structures
in future publications.

It must be emphasized that in all cases the underlying theory is anchored in
2T-physics, and therefore by gauge fixing the Sp$\left(  2,R\right)  $ gauge
symmetry these twistors describe not only massless systems, but much more, as
discussed in this paper and \cite{twistorBP1} with examples. Hence the
twistors play a role in some kind of unification of 1T-physics systems via
dualities, or via higher dimensions with 2T, but in a way that is distinctly
different than the Kaluza-Klein scheme, since there are no Kaluza-Klein
excitations, but instead there is a web of dualities.

In this context it is also interesting that some parameters such as mass,
moduli of some metrics, and some coupling constants for interactions, emerge
from the higher dimensions as moduli while holographically projecting from
$d+2$ dimensions down to $d$ dimensions. Furthermore concepts such as time and
Hamiltonian in 1T-physics are derived concepts that emerge either from
2T-physics and its gauge choices, or from the details of the twistor transform
to 1T-physics systems.

\bigskip{\Large Acknowledgments}\textbf{\bigskip}

I. Bars was supported by the US Department of Energy under grant No.
DE-FG03-84ER40168; M. Pic\'{o}n was supported by the Spanish Ministerio de
Educaci\'{o}n y Ciencia through the grant FIS2005-02761 and EU FEDER funds,
the Generalitat Valenciana and by the EU network MRTN-CT-2004-005104
\textquotedblleft Constituents, Fundamental Forces and Symmetries of the
Universe\textquotedblright. M. Pic\'{o}n wishes to thank the Spanish
Ministerio de Educaci\'{o}n y Ciencia for his FPU research grant, and the USC
Department of Physics and Astronomy for kind hospitality.

\end{document}